\documentclass[sigconf]{acmart}

\AtBeginDocument{%
  \providecommand\BibTeX{{%
    \normalfont B\kern-0.5em{\scshape i\kern-0.25em b}\kern-0.8em\TeX}}}

\setcopyright{acmcopyright}
\copyrightyear{2022}
\acmYear{2022}
\acmDOI{10.1145/3485447.3512264}

\acmConference[WWW '22]{Proceedings of the ACM Web Conference 2022}{April 25--29, 2022}{Virtual Event, Lyon, France}

\acmBooktitle{Proceedings of the ACM Web Conference 2022 (WWW '22), April 25--29, 2022, Virtual Event, Lyon, France}
\acmPrice{15.00}
\acmISBN{978-1-4503-9096-5/22/04}

\usepackage{multirow}
\usepackage{subfigure}
\usepackage{appendix}

\begin{document}

\title[``This is Fake! Shared it by Mistake'': Assessing the Intent of Fake News Spreaders]{``This is Fake! Shared it by Mistake'':\\Assessing the Intent of Fake News Spreaders}

\author{Xinyi Zhou}
\orcid{0000-0002-2388-254X}
\affiliation{
    \institution{Syracuse University}
    \city{Syracuse}
    \state{NY}
    \country{USA}
}
\email{zhouxinyi@data.syr.edu}

\author{Kai Shu}
\affiliation{
    \institution{Illinois Institute of Technology}
    \city{Chicago}
    \state{IL}
    \country{USA}
}
\email{kshu@iit.edu}

\author{Vir V. Phoha}
\affiliation{
    \institution{Syracuse University}
    \city{Syracuse}
    \state{NY}
    \country{USA}
}
\email{vvphoha@syr.edu}

\author{Huan Liu}
\affiliation{
    \institution{Arizona State University}
    \city{Tempe}
    \state{AZ}
    \country{USA}
}
\email{huan.liu@asu.edu}

\author{Reza Zafarani}
\affiliation{
    \institution{Syracuse University}
    \city{Syracuse}
    \state{NY}
    \country{USA}
}
\email{reza@data.syr.edu}

\renewcommand{\shortauthors}{Zhou et al.}

\begin{abstract}
Individuals can be misled by fake news and spread it unintentionally without knowing it is false. This phenomenon has been frequently observed but has not been investigated. Our aim in this work is to assess the intent of fake news spreaders. To distinguish between intentional versus unintentional spreading, we study the psychological explanations of unintentional spreading. With this foundation, we then propose an \textit{influence graph}, using which we assess the intent of fake news spreaders. Our extensive experiments show that the assessed intent can help significantly differentiate between intentional and unintentional fake news spreaders. Furthermore, the estimated intent can significantly improve the current techniques that detect fake news. To our best knowledge, this is the first work to model individuals' intent in fake news spreading. 
\end{abstract}




\maketitle

\section{Introduction}
\label{sec:intro}

\begin{figure}[t]
\begin{minipage}[b]{.475\columnwidth}
    \includegraphics[width=\columnwidth]{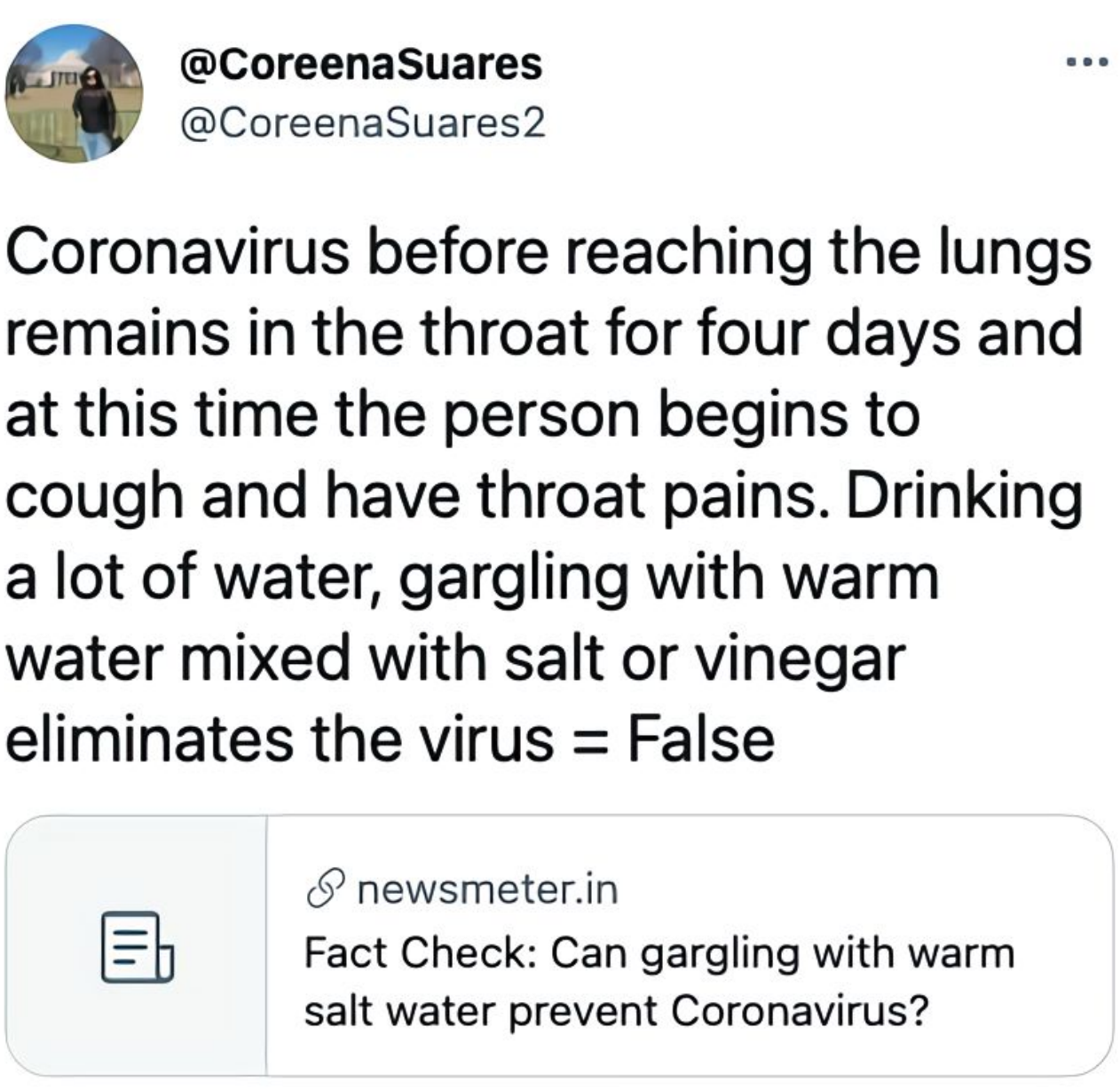}
    \caption{An Example of a Fact-checking Post}
    \label{fig:correction_example}
\end{minipage}\quad
\begin{minipage}[b]{.485\columnwidth}
    \includegraphics[width=\columnwidth]{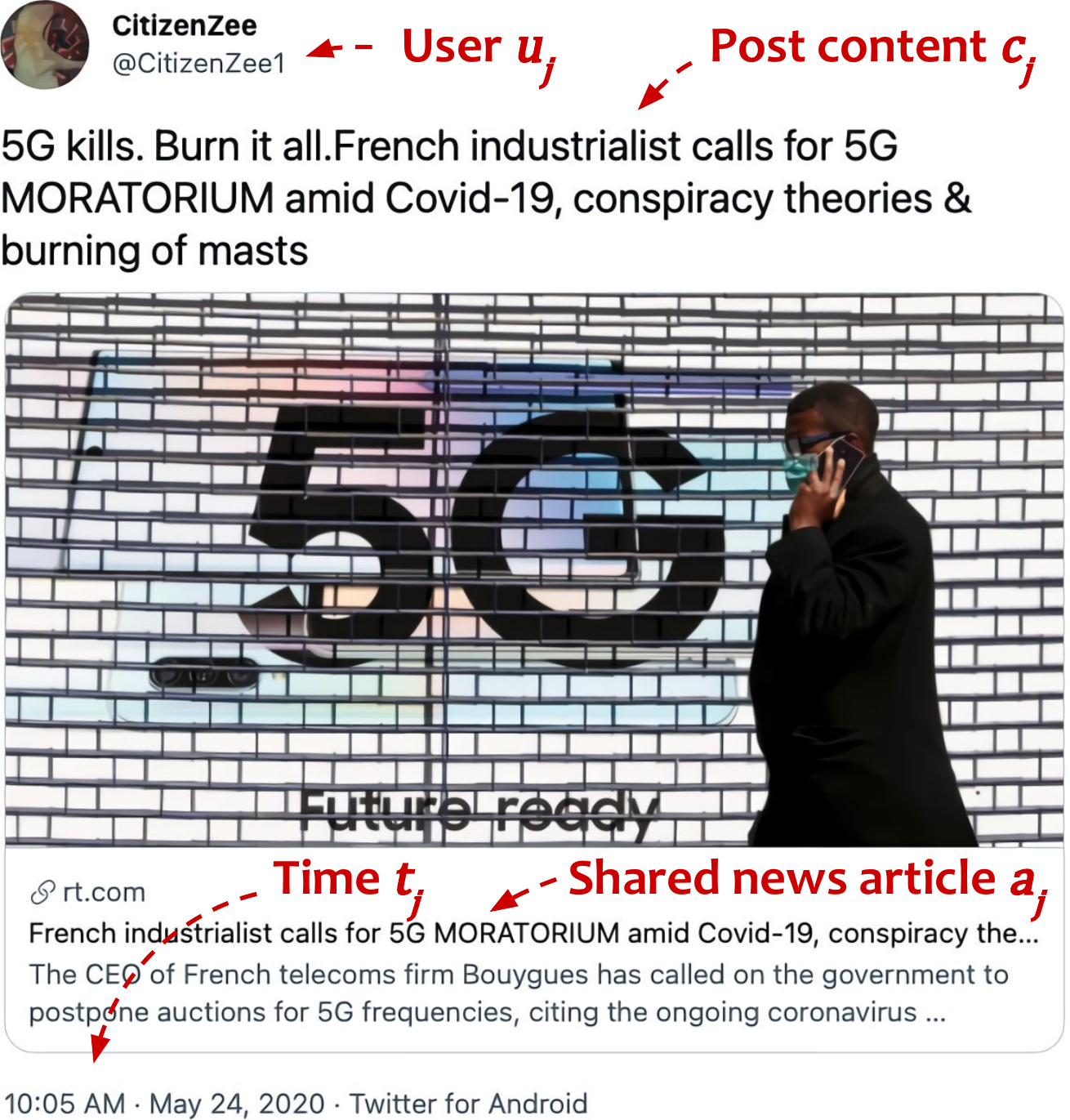}
    \caption{An Illustration of a Post $p_j=(a_j,c_j,t_j,u_j)$}
    \label{fig:post_exaxmple}
\end{minipage}
\end{figure}

A frequently observed and discussed phenomenon is that individuals can be misled by fake news and can unintentionally spread it~\cite{zhou2020survey,pennycook2021shifting,scheufele2019science}. Thankfully, research has pointed out that (1) correction and (2) nudging can effectively prevent such users from spreading fake news. That is, by informing them of the news falsehood, or simply requesting from them to pay attention to news accuracy before spreading the news~\cite{pennycook2021shifting,vo2018rise}. Such findings encourage social media platforms to develop more gentle strategies for these unintentional fake news spreaders to reasonably and effectively combat fake news. Clearly, such strategies should vary from the aggressive deactivation and suspension strategies that platforms adopt for inauthentic or toxic accounts (e.g., Twitter\footnote{\url{https://help.twitter.com/en/rules-and-policies/twitter-rules}} and Facebook\footnote{\url{https://transparency.fb.com/policies/}}). For example, platforms can present such unintentional fake news spreaders with useful facts, motivating the need for new recommendation algorithms. Such algorithms not only recommend topics to these users that they enjoy reading the most (or users they are similar to), but also facts or users active in fact-checking (see Fig.~\ref{fig:correction_example} for an example)~\cite{vo2018rise,margolin2018political,zhou2020survey}.

To determine (1) if correction or nudging is needed for a fake news spreader, (2) whether the spreader should be suspended or deactivated, or (3) which users should be targeted by fact-presenting recommendation algorithms, one needs to assess the \textit{intent} of fake news spreaders. Furthermore, knowing that some users had malicious intent in the past provides a strong signal indicating that their future posts are also potentially fake. This information can be immensely useful for fake news detection~\cite{zhou2020survey}. While determining the intent is extremely important, it is yet to be investigated.

\vspace{1mm}
\noindent \textbf{This work: Assessing Spreading Intent}. We aim to assess the intent of individuals spreading fake news. Our approach to assessing the intent of fake news spreaders relies on fundamental social science theories and exploits advanced machine learning techniques. In particular, we first look into psychological factors that can contribute to the unintentional spreading of fake news (see Section \ref{subsec:social_science}). These factors can be  categorized as \textit{internal influence} and \textit{external influence}~\cite{zhou2020survey}. To capture these factors, and in turn, quantify intent, we propose an \textit{influence graph}; a directed, weighted, and attributed graph. The degree to which fake news spreaders are intentional/unintentional can be assessed with this graph. To evaluate our assessment, we first extend two fake news datasets by introducing annotated intent data of fake news spreaders (\textit{intentional} or \textit{unintentional}) due to the unavailability of ground truth. With this data, we validate the assessed intent and show that it can strongly differentiate between intentional and unintentional fake news spreaders. We further show through experiments that the assessed intent can significantly enhance fake news detection.

The innovation and contribution of this work are:
\begin{enumerate}
    \item \textit{Modeling Fake News Spreading Intent}: To our best knowledge, this is the first work to assess the degree to which fake news spreaders are intentional/unintentional. To this end, we conduct an interdisciplinary study that endows our work with a theoretical foundation and explainability. A new influence graph is proposed that captures factors that contribute to spreading intent as well as multimodal news information.
    
    \item \textit{New Datasets on Intent}: We leverage manual and automatic annotation mechanisms to introduce the ground truth on the intent of fake news spreaders in two large-scale real-world datasets. These are the first two datasets that provide intent information. We conduct extensive experiments using these datasets to validate the assessed intent of fake news spreaders.

    \item \textit{Combating Fake News}: Our work helps combat fake news from two perspectives. First, we demonstrate that by assessing intent, we can successfully distinguish between malicious fake news spreaders (should be blocked) and benign ones (should be presented with facts or nudged). Second, we present the effectiveness of the assessed spreader intent (and the proposed influence graph) in fake news detection.
\end{enumerate}

The rest of the paper is organized as follows. A literature review is first conducted in Section \ref{sec:related_work}. In Section \ref{sec:method}, we specify the method to assess the intent of fake news spreaders, followed by the method evaluation in Section \ref{sec:eval}. We demonstrate the value of assessing intent in combating fake news in Section \ref{sec:method_application}. Finally, we conclude in Section \ref{sec:conclusion} with a discussion on our future work.

\section{Related Work}
\label{sec:related_work}

We first review fundamental social science theories that have been connected to fake news spreading (see Section \ref{subsec:social_science}).
Next, we review the methods developed to combat fake news (see Section \ref{subsec:combat_fake_news}) as we will later utilize the assessed spreader intent to detect fake news.

\subsection{Social Science Foundation of\\Unintentional Fake News Spreading}
\label{subsec:social_science}

Extensive social science research has been conducted on fake news. We particularly review studies that focus on the psychological factors that contribute to the unintentional spreading of fake news.

Lazer et al.~\cite{lazer2018science} attribute this phenomenon to ``individuals prefer information that confirms their preexisting attitudes (\textit{selective exposure}), view information consistent with their preexisting beliefs as more persuasive than dissonant information (\textit{confirmation bias}), and are inclined to accept information that pleases them (\textit{desirability bias}).'' 
Scheufele and Krause~\cite{scheufele2019science} summarize these factors as \textit{confirmation bias}, \textit{selective exposure}, and \textit{motivated reasoning} (i.e., people tend to use emotionally biased reasoning to make most desired decisions rather than those that accurately reflect the evidence). 

Grouping aforementioned psychological factors as an \textit{internal influence}, Zhou and Zafarani~\cite{zhou2020survey} further discuss how the \textit{external influence} on individuals can contribute to their unintentional spreading of fake news. Such social influence can be reflected via, e.g., \textit{availability cascade} (i.e., individuals tend to adopt insights expressed by others when such insights are gaining more popularity)~\cite{kuran1999availability}, \textit{social identity theory}~\cite{ashforth1989social,hogg2020social} (i.e., individuals conform to the behavior of others for being liked and accepted by the community and society), and \textit{validity effect} (e.g., individuals tend to believe information is correct after repeated exposures)~\cite{boehm1994validity,pennycook2018prior}. 

This work shares the social science foundation presented in  \cite{lazer2018science,scheufele2019science,zhou2020survey}. Besides understanding why individuals can be misled by fake news and unintentionally spread it, we further conduct quantitative research to assess user intent. 

\subsection{Methodologies to Combat Fake News}
\label{subsec:combat_fake_news}

The unprecedented growth of fake news and its detrimental impacts on democracies, economies, and public health has increased the demand for automatic methodologies to combat fake news~\cite{zhou2020survey}. With extensive recent contributions by the research community, automatic fake news detection has significantly improved in efficiency and explainability. In general, fake news detection methods can be content-based or propagation-based depending on whether the method focuses on investigating news content or how the news spreads on social media. 

As news articles are mostly text, content-based methods start with manually extracting linguistic features for news representation; LIWC (Linguistic Inquiry and Word Count)~\cite{pennebaker2015development} has been often employed as a comprehensive feature extractor~\cite{rashkin2017truth,perez2018automatic,castelo2019topic}.
Common classifiers, such as SVMs (Support Vector Machines), are then used to predict fake news. With advances in deep learning, recent attention has been paid to employing multimodal (textual and visual) information of news content to detect fake news (see related work such as \cite{wang2018eann,zhou2020multimodal,zhang2020multimodal,qian2021hierarchical,abilov2021voterfraud2020}).
On the other hand, propagation-based methods utilize auxiliary social-media information to predict fake news. Some examples of such information include post stances~\cite{smeros2019scilens}, post-repost relationships~\cite{vosoughi2018spread}, user comments~\cite{shu2019defend}, and profiles~\cite{cheng2021causal}. 

There have been other strategies proposed to combat fake news. For example, education and nudging have been emphasized to improve individuals' ability to recognize misinformation~\cite{lazer2018science,scheufele2019science,lorenz2020behavioural}. Pennycook et al. further provide empirical evidence that unintentional fake news spreading can become less by asking individuals to assess the accuracy of news before attempting to spread it~\cite{pennycook2021shifting}. Lazer et al. suggest incorporating \textit{information quality} into algorithmic rankings or recommendations of online platforms~\cite{lazer2018science}. Studies have also demonstrated that connecting users active in fact checking with fake news spreaders on social networks is an effective way to combat fake news~\cite{vo2018rise,margolin2018political}.

\section{Modeling the Intent of Fake News Spreaders on Social Media}
\label{sec:method}

As presented in Section \ref{subsec:social_science}, psychological factors that contribute to unintentional fake news spreading of individuals can be summarized as two:
(1) internal influence and (2) external influence~\cite{lazer2018science,scheufele2019science,zhou2020survey}. Hence, an individual is more unintentional in spreading a news article if his or her spreading behavior receives more internal and external influence.
Specifically, both \textit{confirmation bias}~\cite{nickerson1998confirmation,mullainathan2005market} and \textit{selective exposure}~\cite{freedman1965selective,metzger2020cognitive} point out that the more consistent an individual's preexisting attitudes and beliefs are with the fake news, the higher the probability that the individual would believe the fake news and unintentionally spread it (internal influence)~\cite{lazer2018science,scheufele2019science,zhou2020survey}. As \textit{availability cascade}~\cite{kuran1999availability} and \textit{social identity theory}~\cite{ashforth1989social,hogg2020social} suggest, individuals can be affected by others as well. An individual would be more unintentional in spreading a fake news article if the spreading follows a \textit{herd behavior}; i.e., the individual's participation matches extensive participation of others and his or her attitude conforms to the attitudes of most participants (external influence)~\cite{zhou2020survey}.

Problems then arise on social media: \textit{where can one find out the preexisting attitudes and beliefs of a user, the participation of users, and their attitudes towards a news article?} We note that a user's preexisting attitudes and beliefs can be reflected in his or her historical activities on social media. For most social media sites, such historical activities include past posts, likes, and comments. Similarly, the participation of users often takes the form of posting, liking, and commenting. Hence, mining the content of posts and comments allows understanding users' attitudes. For simplicity, we start with posts in this work to determine users' preexisting beliefs and participation. In sum, a user spreads a fake news article in his or her post more unintentionally if the post is more similar to or influenced by (1) the user's past posts (internal influence), and (2) the posts of other users (external influence).

A natural approach to capture the influence among posts is to construct an \textit{influence graph} of posts. In this graph, a (directed) edge between two posts indicates the (external or internal) influence flow from one post to the other. The edge weight implies the amount of the influence flow. With this graph, the overall influence that a post receives from other posts can be assessed by looking at its corresponding incoming edges and their weights. The more influence a post that contains fake news receives, the more unintentional is the user who is posting it in spreading this fake news.

To concretely present our proposed influence graph formed by a group of posts, we start by a pair of \textbf{p}osts $p_i$ and $p_j$, which are represented as tuples $(a_i,c_i,t_i,u_i)$ and $(a_j,c_j,t_j,u_j)$, respectively. An example is presented in Fig.~\ref{fig:post_exaxmple}. In the tuple representing post $p_i$, $a_i$ denotes the news \textbf{a}rticle shared by post $p_i$. For simplicity, we first assume that each post can only share one news article (we will consider a more general case later in this section); \textbf{u}ser $u_i$ and \textbf{t}ime $t_i$ refer to the user and posting time of $p_i$; and \textbf{c}ontent $c_i$ is the post content, often containing the attitude and opinion of $u_i$ regarding $a_i$. 
Next, we discuss how (A) internal and (B) external influence between $p_i$ and $p_j$ can be modeled, respectively.\vspace{1mm}

\textit{A. Modeling internal influence between $p_i$ and $p_j$.}
If $p_i$ internally influences $p_j$, $p_i$ should be posted earlier than $p_j$ and by the same user of $p_j$ (to capture preexisting beliefs of the user), i.e., $t_i<t_j$ and $u_i=u_j$. The amount of influence flowing from $p_i$ to $p_j$ can be determined by how similar the news articles and attitudes in $p_j$ are to those of $p_i$. In other words, how similar $a_j$ and $c_j$ are to $a_i$ and $c_i$~\cite{zhou2020survey}. However, evidence has indicated that the same user spreading the same news, especially fake news, is often a sign of intentional spreading rather than unintentional spreading~\cite{shao2018spread}. Therefore, we exclude internal influence from $p_i$ and $p_j$ if $a_i = a_j$.\vspace{1mm}

\textit{B. Modeling external influence between $p_i$ and $p_j$.}
If $p_i$ externally influences $p_j$, $p_i$ should, at least, be posted by a different user from that of $p_j$ (to capture ``external'') and earlier than $p_j$ (otherwise, $p_i$ is not observable to $p_j$); i.e., $t_i<t_j$ and $u_i\neq u_j$. 
We further consider two questions in assessing external influence. 
First, can a user's post spreading one news article externally influence a post of another user spreading a different news article; in other words, if $a_i\neq a_j$, can $p_i$ possibly influences $p_j$ externally with $t_i<t_j$ and $u_i\neq u_j$? Two news articles that differ in text or image may discuss the same event and express the same political stance; hence, this scenario is possible but depends on the similarity between the two news articles~\cite{bessi2015viral}. 
Second, can a user's post possibly be influenced by the other's post if the two users are not socially connected on social media? Due to the platforms' diverse recommendations and services (e.g., the trending in Twitter and Weibo), this scenario is also possible, but the amount of influence depends on how similar news articles and attitudes in $p_j$ are to those of $p_i$~\cite{zhang2017users,zhou2020survey}.\vspace{1mm}

We summarize the above discussions by answering the following three questions:

\begin{figure}[t]
    \subfigure[Without (upper fig.) v.s. With (down) Influence]{\label{subfig:a}
        \includegraphics[width=0.28\columnwidth]{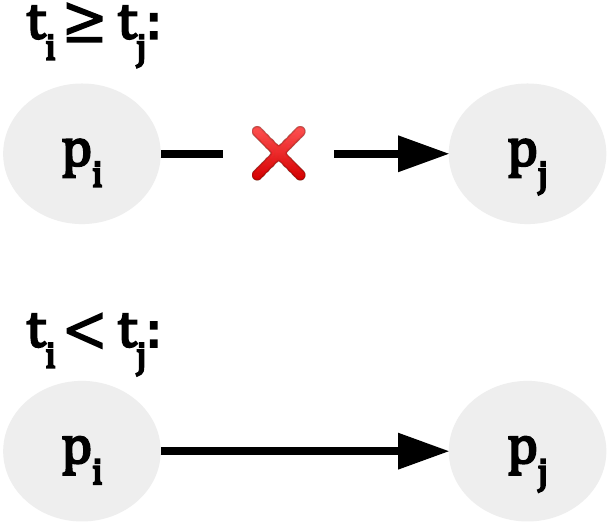}
    }\quad
    \subfigure[External (up) v.s. Internal (down) Influence]{\label{subfig:c}
        \includegraphics[width=0.28\columnwidth]{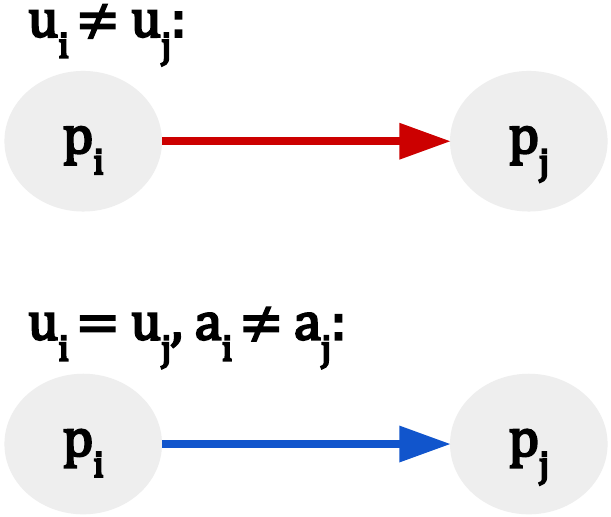}
    }\quad
    \subfigure[Large (up) v.s. Small (down) Volume of Influence]{\label{subfig:b}
        \includegraphics[width=0.3\columnwidth]{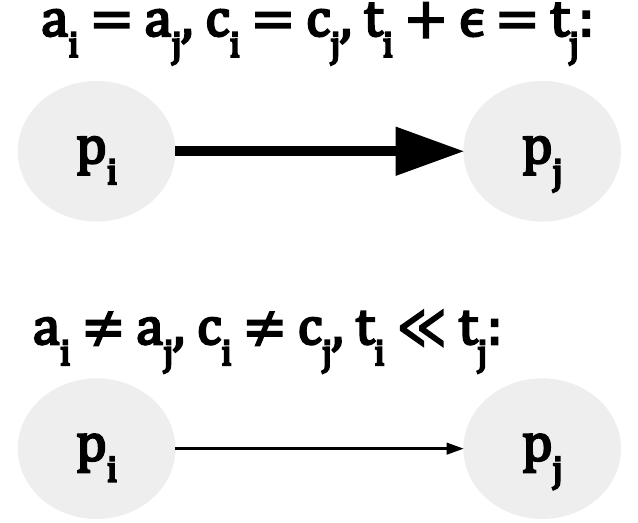}
    }
    \label{fig:pairwise_post_influence}
    \caption{Pairwise Influence of Posts $p_i$ and $p_j$: (a) decides if there is an edge from $p_i$ to $p_j$ in an influence graph; (b) identifies the edge attribute; and (c) determines the edge weight.}
\end{figure}

\begin{enumerate}
    \item \textit{Edge existence: Can $p_i$ possibly influence $p_j$?} As discussed, it is barely possible for $p_i$ to (internally or externally) affect $p_j$, if it is posted later than $p_j$. Hence, in an influence graph, a directed edge can possibly exist from $p_i$ to $p_j$, if $p_i$ is posted earlier than $p_j$ (i.e., $t_i<t_j$); if $t_i\geq t_j$, no edge exists from $p_i$ to $p_j$. Therefore, there can be either no edge or only one directed edge between two posts. See Fig.~\ref{subfig:a} for an illustration. Note that whether an edge ultimately exists between two posts also depends on the edge weight (we specify below in 3); a zero weight can make an edge ``disappear.''

    \item \textit{Edge attribute: What type of influence (internal vs external) is flowing between $p_i$ and $p_j$?} We define the influence as external, if $p_i$ and $p_j$ are posted by different users, i.e., $u_i\neq u_j$~\cite{zhang2017users}. The influence is internal, if $p_i$ and $p_j$ are posted by the same user and do not share the same news, i.e., $u_i= u_j$ and $a_i\neq a_j$~\cite{shao2018spread}. See Fig.~\ref{subfig:c} for an illustration.

    \item \textit{Edge weight: How much influence flows from $p_i$ to $p_j$?} We assume that the amount of influence flow is affected by three factors. The first, as discussed, is the \textit{news articles shared by $p_i$ and $p_j$ ($a_i$ versus $a_j$)}; basically, if $p_i$ and $p_j$ spread the same news, the influence flow between them should be greater compared to if they spread completely different news articles~\cite{bessi2015viral,zhou2020survey}. The second, as discussed, is the \textit{attitudes held by $p_i$ and $p_j$ on the news ($c_i$ versus $c_j$)}; basically, if two posts agree with each other, the influence flow between them should be greater compared to if they disagree with each other~\cite{zhou2020survey}. Furthermore, we consider the \textit{time interval between $p_i$ and $p_j$ ($t_i$ versus $t_j$)}; instead of ``remembering all'', users forget past news articles and their corresponding posts over time (with some decay)~\cite{wozniak1995two}. Thus, a greater amount of influence flow is assigned to two posts when one is published close in time to the other, compared to those that are published farther apart. See Fig.~\ref{subfig:b} for an illustration.
\end{enumerate}

Next, we formalize the proposed \textit{influence graph} (see Definition \ref{def:inf_net}), and introduce how the intent of (fake) news spreaders can be quantified based on this graph. Clearly, in a real-world scenario, it is possible for a post to contain more than one news article (e.g., multiple URLs). Hence, in this formalization, we no longer assume that each post can only share one news article and generalize to a set of articles, i.e., $(a_i,c_i,t_i,u_i)$ becomes $(A_i,c_i,t_i,u_i)$. 

\begin{definition}[Influence Graph]
\label{def:inf_net}
Given a set of news articles, denoted as $A = \{a_1, a_2, \cdots, a_m\}$, we denote user posts that share these news articles on social media as $P = \{p_1, p_2, \cdots, p_n\}$. Each post $p_i$ $(i=1,2,\cdots,n)$ is represented as a tuple $(A_i,c_i,t_i,u_i)$, where $A_i$, $c_i$, $t_i$, and $u_i$ respectively refer to a set of news articles (can be one article) shared by the post (i.e., $A_i \subseteq A$), the post content, the posting time, and the user.

Influence graph, denoted as $G=(V,E,\mathbf{W})$, is formed by user posts, i.e., $V=P$. 
Edges exist from $p_i$ to $p_j$ if (i) $p_i$ is posted earlier than $p_j$, and (ii) $p_i$ and $p_j$ do not share the same news when posted by the same user. In other words, $(p_i, p_j)\in E$ if (i) $t_i<t_j$ and (ii) $A_i\neq A_j$ for $u_i= u_j$.
The edge weight for $(p_i, p_j)$ is
\begin{equation}
    \mathbf{W}_{ij} = \bar{\mathcal{S}}(A_i,A_j)\cdot \mathcal{S}(c_i,c_j)\cdot \mathcal{T}(\Delta t_{ij}),
\end{equation}
where $\mathcal{S}(*_i,*_j)$ assesses the similarity between $*_i$ and $*_j$, $\mathcal{T}(\Delta t_{ij})$ for $\Delta t_{ij}=t_j-t_i$ is a self-defined monotonically decreasing decay function to capture users' forgetting, and
$\bar{\mathcal{S}}(A_i,A_j)$ computes the average pairwise similarity among news pairs $(a_i,a_j)\in A_i\times A_j$. Formally,
\begin{equation}
    \bar{\mathcal{S}}(A_i,A_j) = \frac{\sum_{(a_k,a_l)\in A_i\times A_j}\mathcal{S}(a_k,a_l)}{|A_i|\times |A_j|};
\end{equation}
hence, $\bar{\mathcal{S}}(A_i,A_j)=\mathcal{S}(a_k,a_l)$ if $A_i=\{a_k\}$ and $A_j=\{a_l\}$.
\end{definition}

Based on the above graph, the overall influence on each post, which we denote as the \textit{affected degree}, is computed as
\begin{equation}
\label{eq:overall_affected_degree}
    \mathbf{f}_j = \sum_{(p_i,p_j)\in E} \mathbf{W}_{ij},
\end{equation}
where the external and internal influence, respectively, refer to
\begin{equation}
\begin{array} {llll}
\label{eq:affected_degrees}
\mathbf{f}^{\textsc{External}}_j & = & \sum_{(p_i,p_j)\in E} \mathbf{W}_{ij} & \text{if}~u_i \neq u_j; \\
\mathbf{f}^{\textsc{Internal}}_j & = & \sum_{(p_i,p_j)\in E} \mathbf{W}_{ij} & \text{if}~u_i = u_j.
\end{array}
\end{equation}

For posts sharing fake news articles,  greater values of $\mathbf{f}^{\textsc{External}}_j$, $\mathbf{f}^{\textsc{Internal}}_j$, and $\mathbf{f}_j$ indicate that user $j$ receives more external, internal, and combined (external+internal) influence when spreading the fake news article, i.e., the user engages more unintentionally. Conversely, smaller values of $\mathbf{f}^{\textsc{External}}_j$, $\mathbf{f}^{\textsc{Internal}}_j$, and $\mathbf{f}_j$ indicate that the user is affected less and engages more intentionally in fake news spreading.\footnote{The statement also holds for posts sharing true news articles.} 

\vspace{0.5em}
\noindent \textbf{Customized Implementation Details.}
The implementation of influence graph has several customizable parts; it can be modified by defining different $\mathcal{T}$,  developing different techniques to represent news articles and user posts, and designing ways to compute their similarities. Below are our implementations and justifications.

To represent news articles and posts, we investigate both textual and visual information within the content. 
Textual features are extracted using Transformers, which have excellently performed in understanding semantics of text and various NLP (Natural Language Processing) tasks such as machine translation and sentiment analysis~\cite{vaswani2017attention,yin2020sentibert}. As user posts are often short and within 512 words (e.g., on Twitter, the number of words are not allowed to exceed 280),\footnote{\url{https://developer.twitter.com/en/docs/counting-characters}} we use a pre-trained Sentence-RoBERTa model, which modifies RoBERTa by the Siamese network, to obtain the post embedding~\cite{reimers2019sentence}; the model performs best in the task of semantic textual similarity.\footnote{\url{https://github.com/UKPLab/sentence-transformers}} Differently, as news articles are often long and over 512 words,\footnote{As \cite{zhou2020dataset} suggests, the number of words of news articles published by mainstream and fake news medium has a mean value around 800 and median value around 600.} we employ Longformer~\cite{beltagy2020longformer} to derive the semantically meaningful text embedding of news articles. Longformer addresses the limitation of 512 tokens in BERT by reducing the quadratic scaling (with the input sequence) to linear~\cite{beltagy2020longformer}. For visual features, we extract them using a pre-trained DeepRanking model particularly designed for the task of fine-grained image similarity computation~\cite{wang2014learning}. 
With textual features of news (or post) denoted as $\mathbf{t}$, and its visual features denoted as $\mathbf{v}$, we define the similarity between a news (or post) pair as 
\begin{equation}
    \mathcal{S}(*_i, *_j) = \mu \hat{\cos} (\mathbf{t}_{*_i}, \mathbf{t}_{*_j}) + (1-\mu) \hat{\cos} (\mathbf{v}_{*_i}, \mathbf{v}_{*_j}),
\end{equation}
where $*=a~\text{(for news)}$ or $p~\text{(for posts)}$; $\hat{\cos}(.,.)=[1-\cos(.,.)]/2$; and $\mu$, $\hat{\cos}(.,.)$, $\mathcal{S}(.,.)\in [0,1]$. In our experiments, we determine the value of $\mu$ by varying it from 0.1 to 0.9 with a step size 0.1; we set $\mu=0.8$ that leads to the best evaluation and prediction results.

As for decay function $\mathcal{T}$, we define it as \begin{equation}
    \mathcal{T}(\Delta t_{ij}) = e^{1 - \Delta t_{ij}},
\end{equation}
which is inspired by \cite{wozniak1995two}. $\Delta t_{ij} = t_j-t_i$ and $t_i$ indicates the chronological ranking of post $p_i$ (i.e., $t_i\in \mathbb{Z}^+$); hence, $\mathcal{T}(.)\in (0,1]$ due to $t_j>t_i$. 
The benefit of such $\mathcal{T}$ is two fold. First, it
helps normalize the affected degree for any influence graph. Specifically, let $\mathbf{f}^*_j$ denote either of $\mathbf{f}_j$, $\mathbf{f}^{\textsc{Internal}}_j$, or $\mathbf{f}^{\textsc{External}}_j$. Let $\hat{\mathbf{f}}^*_j$ denote the normalized version of $\mathbf{f}^*_j$, i.e., $\hat{\mathbf{f}}^*_j\in [0,1]$ (accurately, here $\hat{\mathbf{f}}^*_j\in [0,1)$). Then, for $\mathbf{f}^*_j$ we have 
\begin{equation}
\begin{array}{lll}
\mathbf{f}^*_j & =    & \sum_{(p_i,p_j)\in E} \mathcal{S}(A_i,A_j)\cdot \mathcal{S}(c_i,c_j)\cdot \mathcal{T} (\Delta t_{ij}) \\
    & \leq & \sum_{(p_i,p_j)\in E} \mathcal{T} (\Delta t_{ij}) \\
    & < & \sum_{k = 1}^{\infty} e^{1 - k} \\
    & =    & e(e-1)^{-1}.
\end{array}
\end{equation}
In other words, the upper bound of the affected degree, denoted by $f_{\max}$, is $e(e-1)^{-1}$. 
Strictly speaking, $K$ posts ($K>1$) can be posted at the same time in a real-world scenario, i.e., their ranking, denoted by $t_X$, is the same.
We point out that the upper bound $f_{\max}$ still holds in this case, if the ranking value after $t_X$ is $t_{X}+K$ rather than $t_{X}+1$.
Finally, the normalized affected degree $\hat{\mathbf{f}}^*_j$ for post $p_j$ is 
\begin{equation}
\label{eq:normalized_affected_degree}
\hat{\mathbf{f}}^*_j = \frac{1}{f_{\max}}\mathbf{f}^*_j =  \frac{e-1}{e}\mathbf{f}^*_j. 
\end{equation} 
Secondly, in the worst case, influence graph can be a \textit{tournament}, taking up much space. Such $\mathcal{T}$ facilitates \textit{graph sparsification}, while maintaining the performance on tasks (see details in Appendix \ref{apx:graph_sparsification}). Lastly, we note that we have tested $\Delta t_{ij}$ (the time interval) with various units (seconds/minutes/hours/days) in addition to chronological rankings; still, the ranking performs best in all experiments.

\section{Method Evaluation}
\label{sec:eval}

In this section, we evaluate the proposed method in assessing the intent of fake news spreaders. To this end, evaluation data is required that contains the ground-truth label on
\begin{itemize}
    \item \textit{News credibility}, i.e., whether a news article is fake news or true news; and
    \item \textit{Spreader intent}, i.e., whether a user spreads a fake news article intentionally or unintentionally on social media.
\end{itemize} 
We point out that this work is the first to model individuals' intent in fake news propagation. Therefore, no data exists that contains the ground-truth label on spreader intent, let alone both news credibility and spreader intent. Next, we first detail how this problem is addressed in Section \ref{subsec:dataset}, followed by the method evaluation results in Section \ref{subsec:evaluation}.

\subsection{Datasets and Annotations}
\label{subsec:dataset}

Our experiments to evaluate the proposed method are based on two datasets developed for news credibility research: MM-COVID~\cite{li2020mmcovid} and ReCOVery~\cite{zhou2020dataset}. Generally speaking, both datasets collect news information verified by domain experts (labeled as \textit{true} or \textit{fake}) and how the news spreads on Twitter. The corresponding data statistics are in Tab.~\ref{subtab:news_credibility} in Appendix \ref{apx:data_statistics}; we focus on the news with social context information, and on the English news and tweets to which all pre-trained models can be applied.

Although the ground-truth label on news credibility is available, both datasets do not provide annotations on intent of fake news spreaders. We first consider \textit{manual annotation} to address this problem. Specifically, we invite one expert knowledgeable in misinformation area and one graduate student generally aware of the area. We randomly sample 300 posts (unique in tweet ID and user ID) from MM-COVID and ReCOVery that contain fake news (i.e., users of these posts are all fake news spreaders). Before annotating, we first inform the annotators with the definition and general characteristics of unintentional fake news spreaders. That is, as presented in Section \ref{sec:intro}: these spreaders are misled by fake news, barely recognize it is fake, tend to believe in the fake news; meanwhile, if informed on news falsehood or presented with facts, such spreading behavior of them can be reduced, or even stopped. In annotating, we present the two annotators with 
\begin{itemize}
    \item The tweet's link that spreads fake news, which allows annotators to access the tweet details (as illustrated in Fig.~\ref{fig:post_exaxmple}).
    \item The user's link who posts the tweet, which allows annotators to access the user's profile and historical activities. 
\end{itemize}
For each post, we ask the two annotators to 
\begin{enumerate}
    \item Annotate if the user spreads the fake news unintentionally (with an answer of \textit{yes} or \textit{no});
    \item Present the confidence level (detailed below);
    \item Explain the annotation with evidence; and
    \item Provide an estimate on the time spent on annotation. 
\end{enumerate}
We provide three optional levels of confidence. 0 indicates the annotation result is a random guess; no evidence is found to help annotation, or half the evidence supports but the other half rejects the annotation result. 0.5 indicates a medium-level confidence; among all the evidence that the annotator finds, some of them reject but most of them support the annotation result. 1 indicates a high-level confidence; all the evidence that the annotator finds support the annotation result. 

With the returned annotations, we compute the agreement of the two annotators by Cohen's $\kappa$ coefficient~\cite{cohen1960coefficient}. $\kappa=0.61$, removing annotations with no confidence; in other words, two annotators substantially agree with each other~\cite{cohen1960coefficient}. To further obtain the ground truth, we only consider the annotations with a confidence score $\geq 0.5$ and agreed by the two annotators. Finally, 119 posts sharing fake news have the ground-truth label on their users' intent, among which 59 are unintentional and 60 are intentional.

We point out that annotating intent of fake news spreaders is a time-consuming and challenging task. Around five minutes is required to annotate each instance on average. Understanding the user intent behind a post demands evaluating the tweet content and studying the user based on his or her historical behavior on social media. Such manual annotation for large-scale data is hence impractical, which drives us to consider \textit{algorithmic annotation} that accurately simulates manual annotation in an automatic manner. 
Interestingly, we observe that annotators are more confident in identifying intentional fake news spreaders than unintentional ones. Specifically, the expert annotator is at 0.93 confidence level in identifying intentional fake news spreaders and at 0.75 confidence level in identifying unintentional fake news spreaders. For the graduate student annotator, the confidence score is 0.84 and 0.57, respectively. Both results have $p\ll 0.001$ with Mann–Whitney U test. To conduct algorithmic annotation that can accurately simulate manual annotation, we thus start to think \textit{``what kind of fake news spreaders can be intentional?"}

With the explanations given by annotators, we can reasonably assume bots and trolls who have engaged in fake news propagation as intentional fake news spreaders. As inauthentic and toxic accounts, bots and trolls have been often suspended or deactivated by social media platforms (e.g., Twitter and Facebook) regardless of spreading fake news or not. In fact, they have played a significant role in fake news dissemination~\cite{shao2018spread,starbird2019disinformation,zhou2020survey,ferrara2020types}. As a comparison, unintentional fake news spreaders deserve a ``gentle'' strategy developed by social media platforms: nudging and fact-presenting recommendation are more reasonable than suspension and deactivation, as we specified in Section \ref{sec:intro}.
Therefore, we separate bots and trolls from unintentional fake news spreaders. We further notice that users active in fact-checking can spread fake news as well, in a \textit{correction} manner; i.e., they clarify news is false (objectively, and not aggressively) and inform other users of it in their spreading. We call the corresponding posts that spread fake news \textit{correction posts} and these users \textit{correctors} later in the paper. These correctors enables recognizing news falsehood. We thus separate them from unintentional fake news spreaders.

We identify bots and trolls by collecting data from two well-established and widely accepted platforms, Botometer~\cite{sayyadiharikandeh2020detection}\footnote{\url{https://botometer.osome.iu.edu/}\label{ft:botometer}} and Bot Sentinel.\footnote{\url{https://botsentinel.com/}\label{ft:botsentinel}} Ultimately, each Twitter user is assigned a bot score (denoted as $b$) and a troll score (denoted as $r$), where $b,c\in[0, 1]$. To identify correctors, we first annotate each tweet as a correction or non-correction tweet. Then, we assign each fake news spreader a corrector score (denoted as $c$, where $c\in[0,1]$) by computing the proportion of the user's correction tweets to his or her total tweets that share fake news. With a threshold value, $\theta\in[0,1]$, each fake news spreader can be classified as (i) bot (if $b\in[0,\theta)$) or non-bot (if $b\in[\theta,1]$), (ii) troll (if $r\in[0,\theta)$) or non-troll (if $r\in[\theta,1]$), and (iii) corrector (if $c\in[0,\theta)$) or non-corrector (if $c\in[\theta,1]$).

With identified bots, trolls, and correctors (here, we use 0.5 as the threshold, i.e., $\theta=0.5$), the algorithmic annotation on intent of fake news spreaders is conducted at two levels: (i) tweet-level and (ii) user-level. At the tweet-level, the algorithm labels all correction tweets and tweets of bots and trolls that share fake news as intentional spreading. The tweet-level annotation captures the user intent for \textit{each} spreading action of fake news. At the user-level, the algorithm labels all bots, trolls, and correctors as intentional spreaders. The user-level annotation captures the \textit{general} user intent when spreading fake news. Tab.~\ref{subtab:user_intention} in Appendix \ref{apx:data_statistics} summarizes the corresponding data statistics.

\vspace{1mm}
\noindent \textbf{Evaluating Algorithmic Annotations.} We compare the algorithmic annotation results with the manual annotations. Results are shown in Tab.~\ref{tab:alg_ann_eval}; results are the same at both the tweet- and user-levels. We observe that the algorithmic annotation effectively simulates the manual annotation, whose AUC score is above 0.8 using sampled MM-COVID and/or ReCOVery datasets. Automatic and manual annotations have a substantial agreement with Cohen's $\kappa$ coefficient above 0.64~\cite{cohen1960coefficient}. 

\begin{table}[t]
\centering
\caption{Performance of Algorithmic Annotations on Intent of Fake News Spreaders}
\label{tab:alg_ann_eval}
\begin{tabular}{lrrr}
\toprule[1pt]
      &   \textbf{AUC Score}    & \textbf{Cohen's $\kappa$} \\ \midrule[0.5pt]
\textbf{MM-COVID + ReCOVery}     &   0.8824          & 0.7482                    \\ 
\textbf{MM-COVID}              &   0.8857          & 0.7520                    \\ 
\textbf{ReCOVery}              &   0.8000          & 0.6484                    \\ \bottomrule[1pt]
\end{tabular}
\end{table}

\begin{figure*}[t]
\begin{minipage}{.48\textwidth}
    \subfigure[MM-COVID ($p\ll0.001$ with t-test)]{
    \includegraphics[width=0.46\columnwidth]{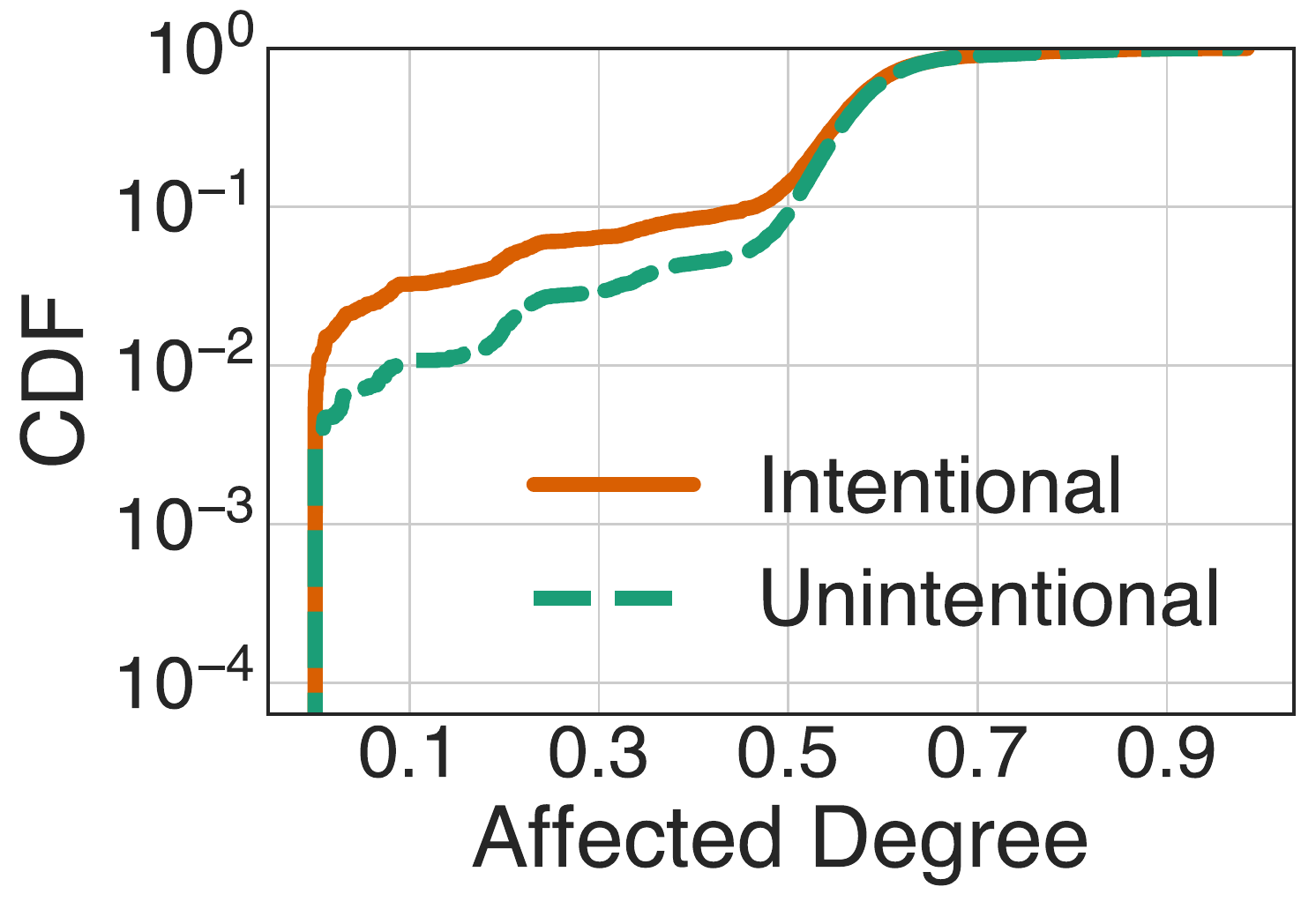}}\quad
    \subfigure[ReCOVery ($p<0.01$ with t-test)]{
    \includegraphics[width=0.46\columnwidth]{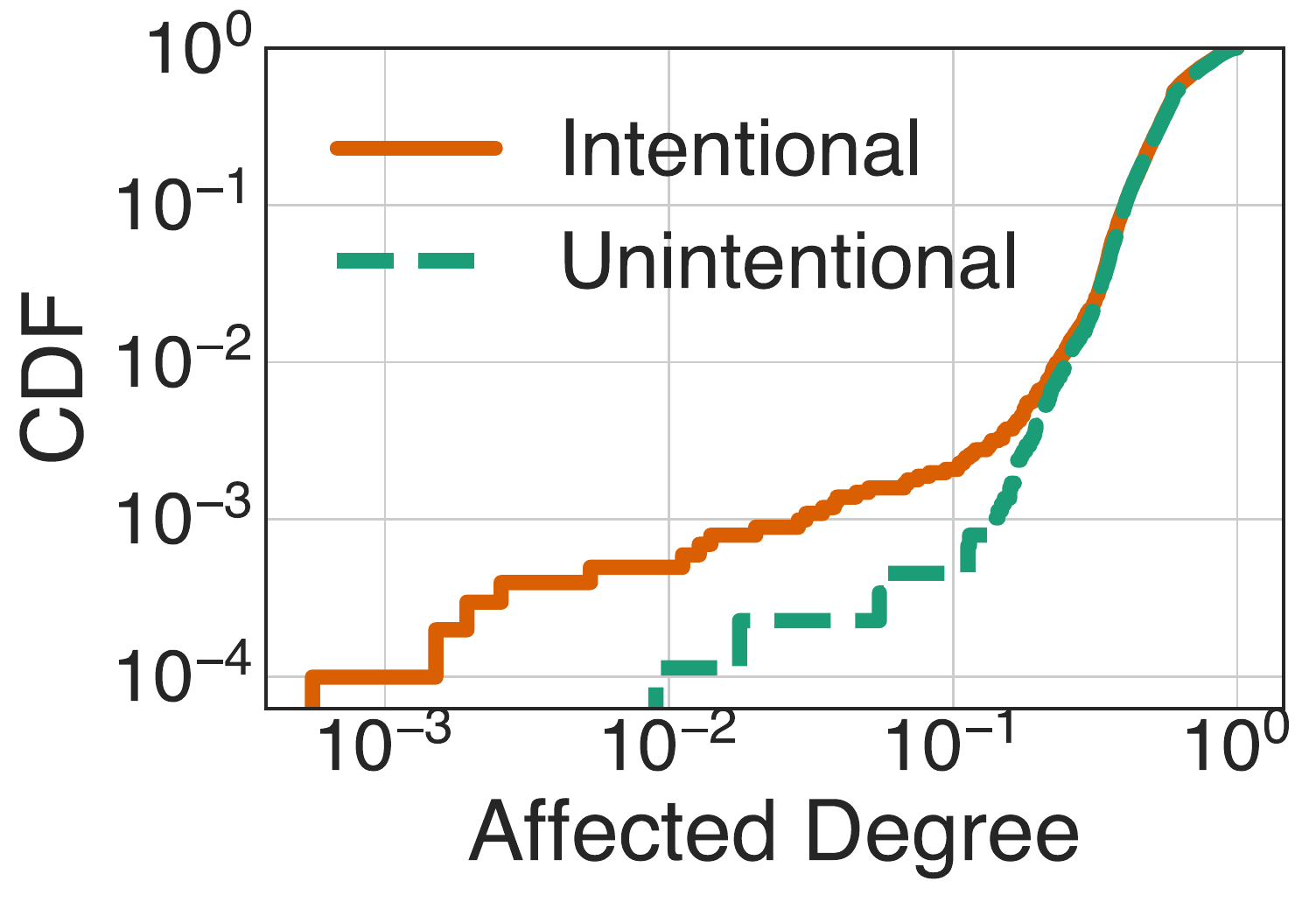}}
    \caption{Distribution of Affected Degree: Intentional Fake News Spreaders v.s. Unintentional Fake News Spreaders}
    \label{fig:influence_vs_intentional_cdf}
\end{minipage}\qquad
\begin{minipage}{.48\textwidth}
    \subfigure[MM-COVID ($p\ll0.001$ by ANOVA)]{
    \includegraphics[width=0.48\columnwidth]{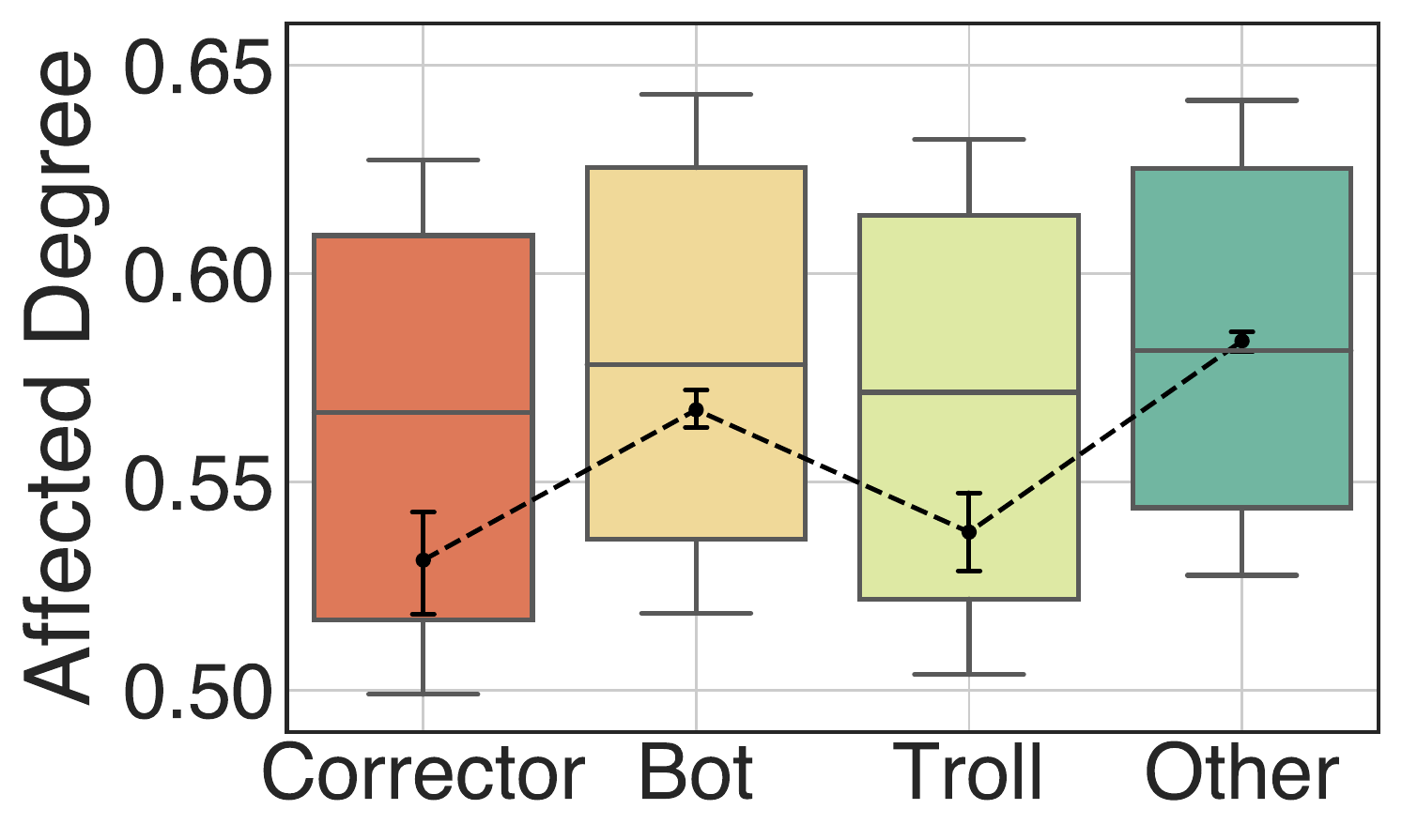}}
    \subfigure[ReCOVery ($p<0.01$ by ANOVA)]{
    \includegraphics[width=0.47\columnwidth]{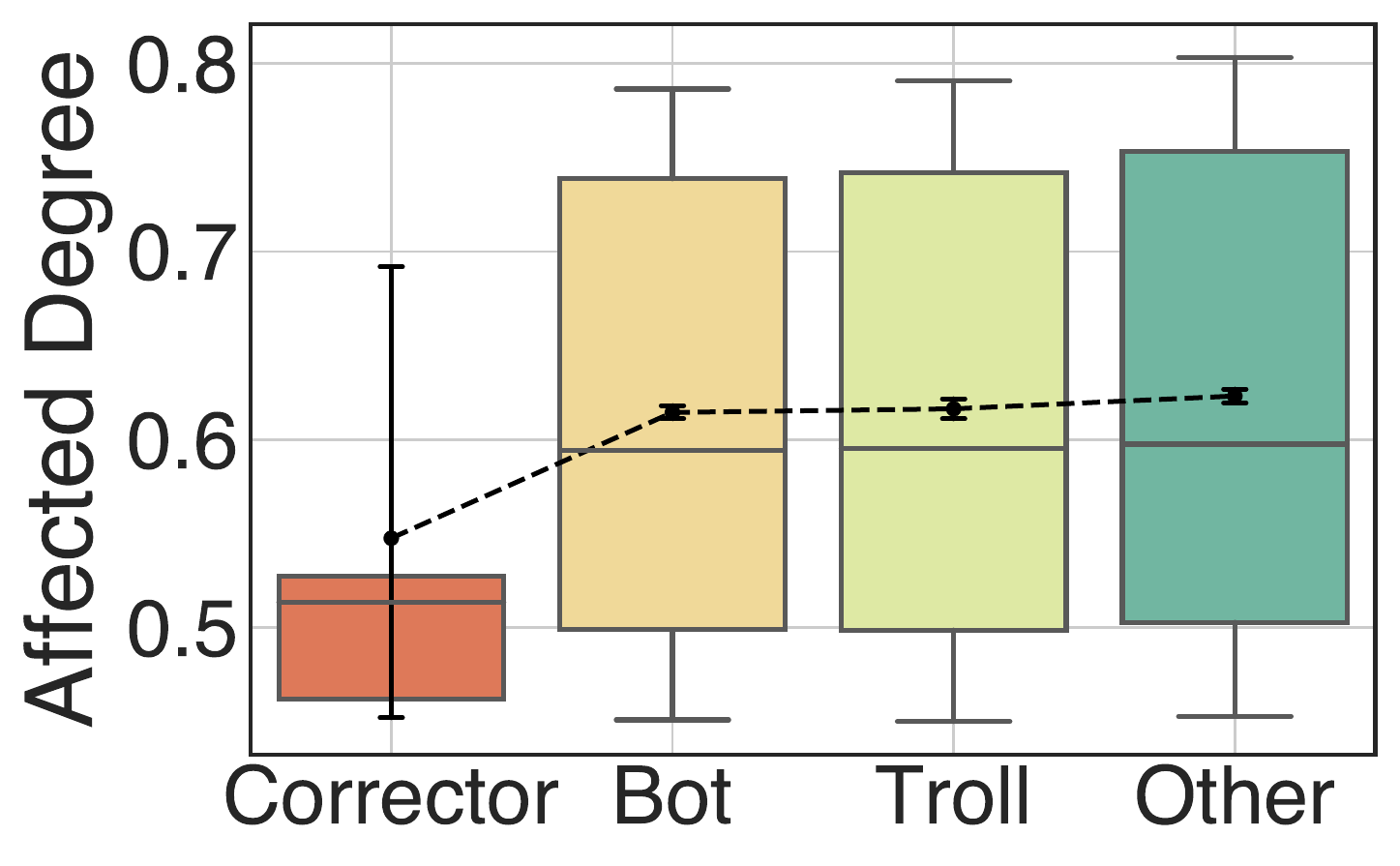}}
    \caption{Affected Degree of Bots, Trolls, Correctors, and Others (First Three: Intentional Fake News Spreaders; Others: Unintentional Fake News Spreaders)}
    \label{fig:influence_vs_intentional_box}
\end{minipage}

\begin{minipage}{.48\textwidth}
    \subfigure[MM-COVID ($p\ll0.001$ using t-test for the right)]{
        \includegraphics[width=0.32\columnwidth]{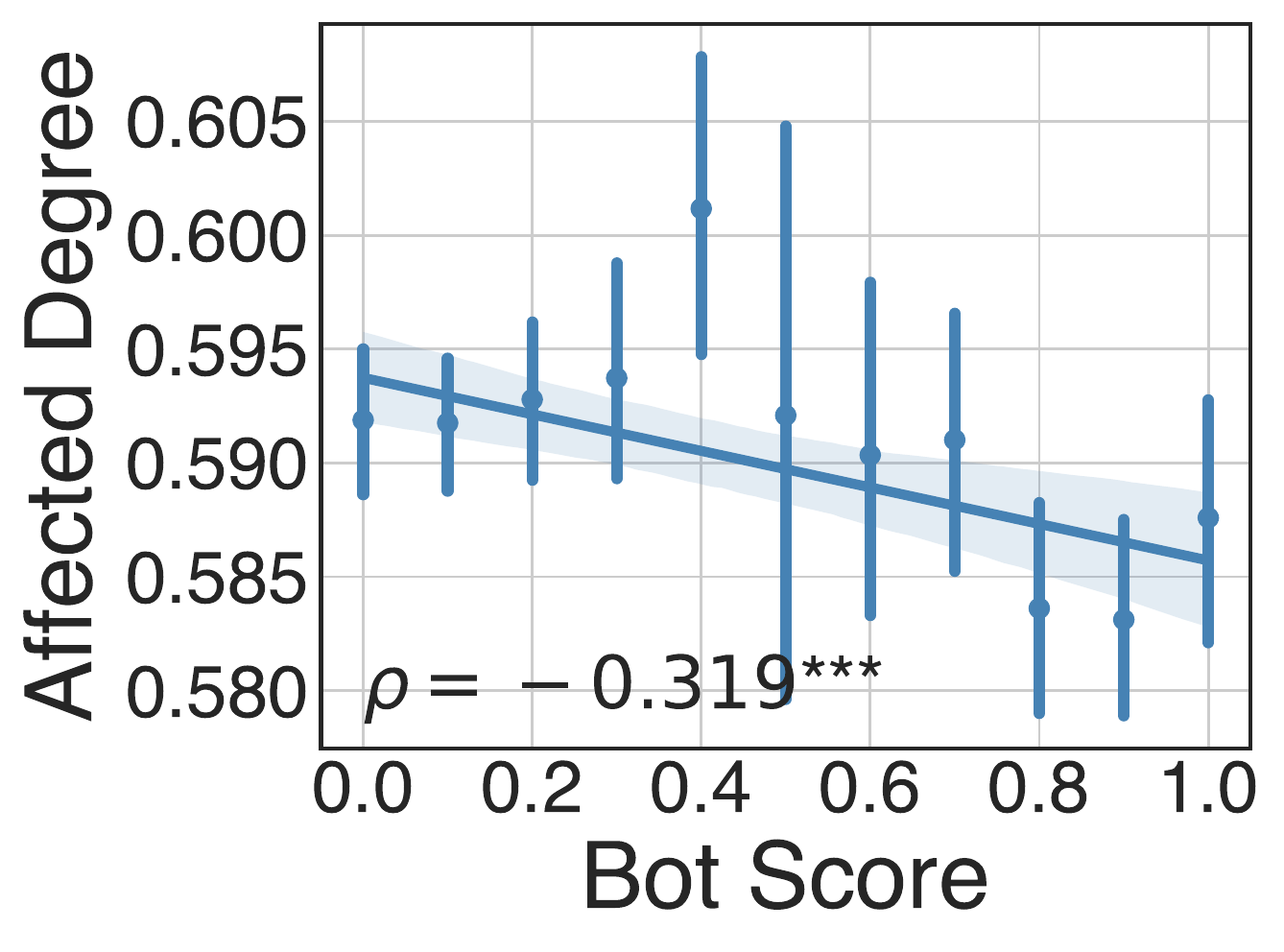}
        \includegraphics[width=0.32\columnwidth]{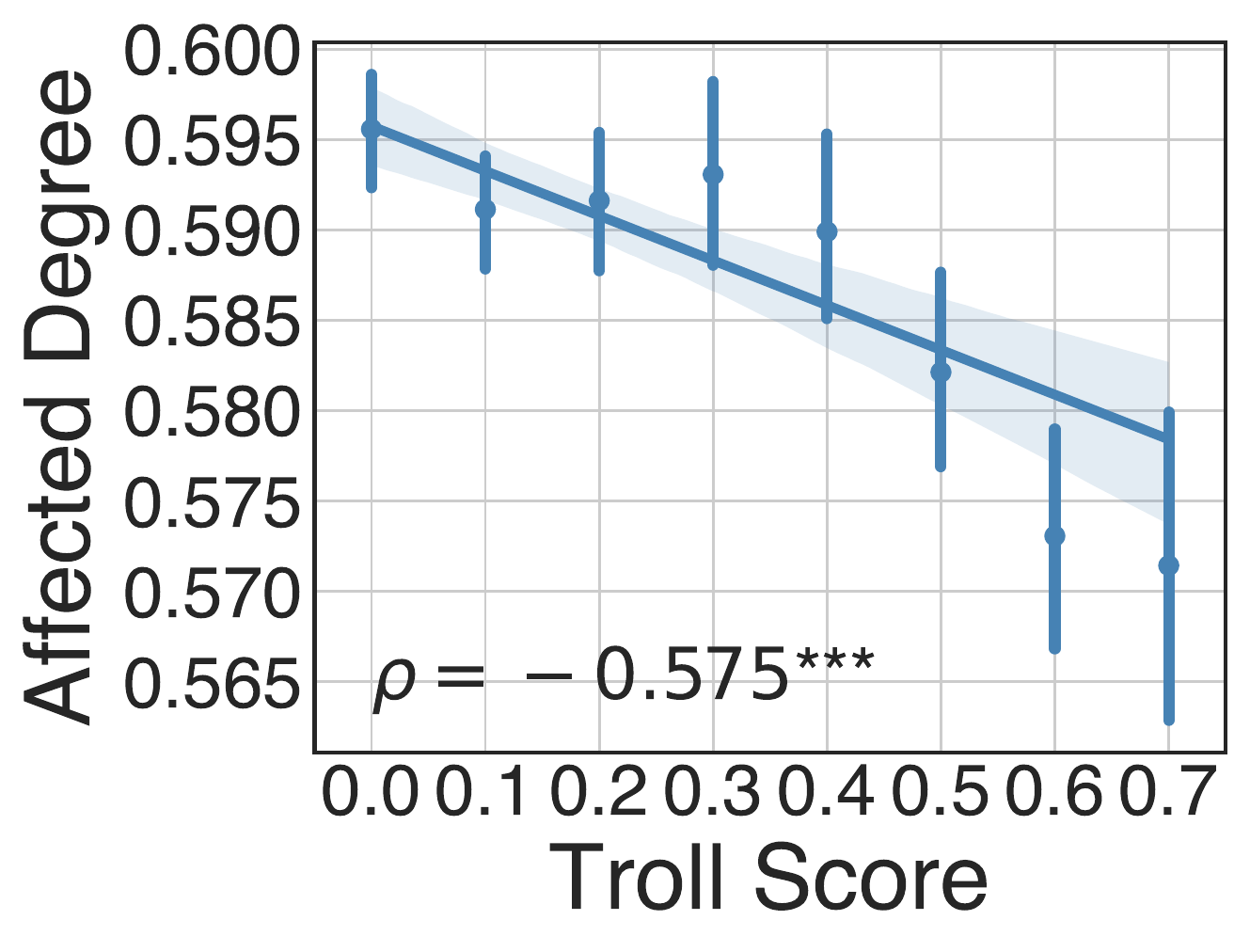}
        \includegraphics[width=0.32\columnwidth]{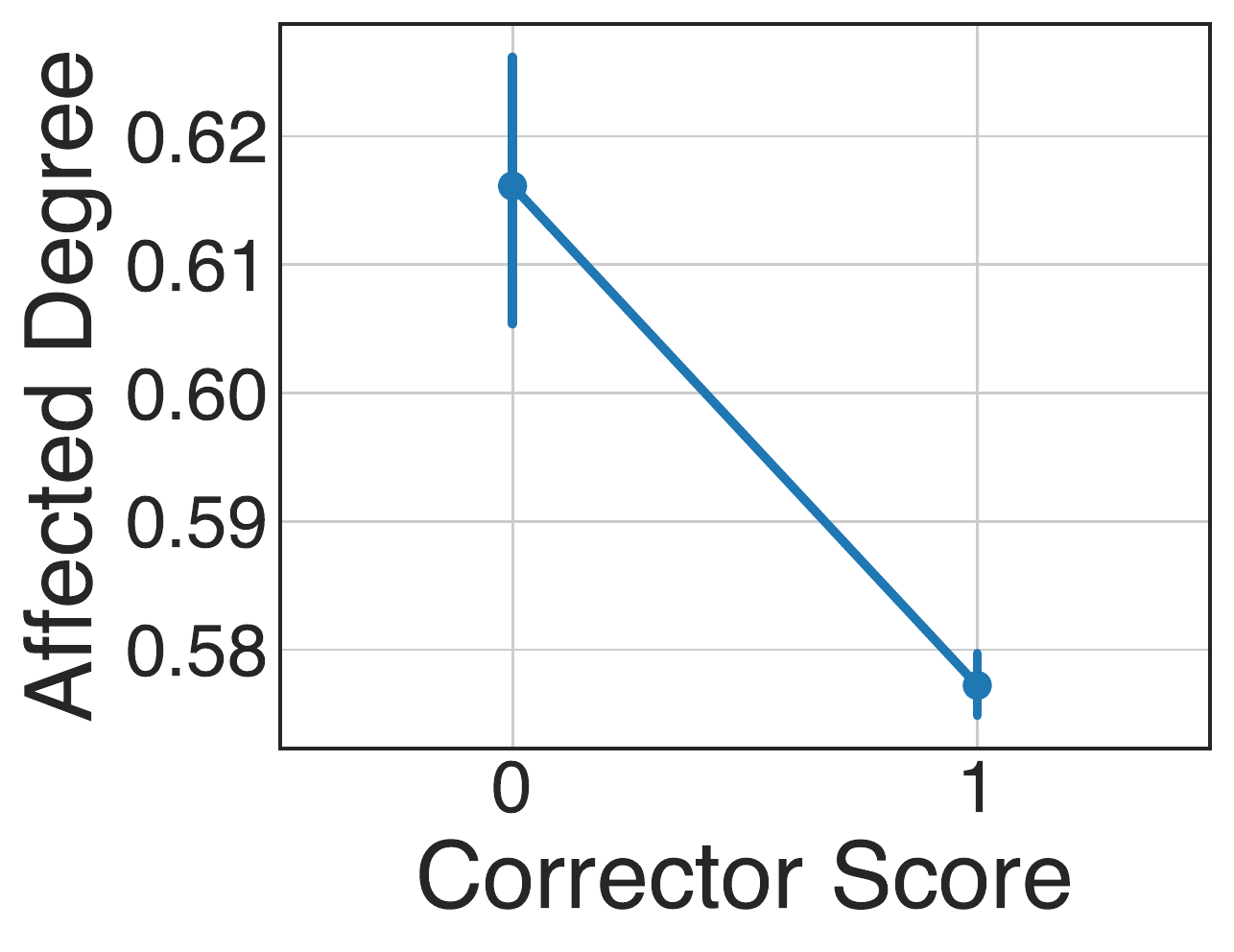}}

    \subfigure[ReCOVery ($p\ll0.001$ using t-test for the right)]{
        \includegraphics[width=0.32\columnwidth]{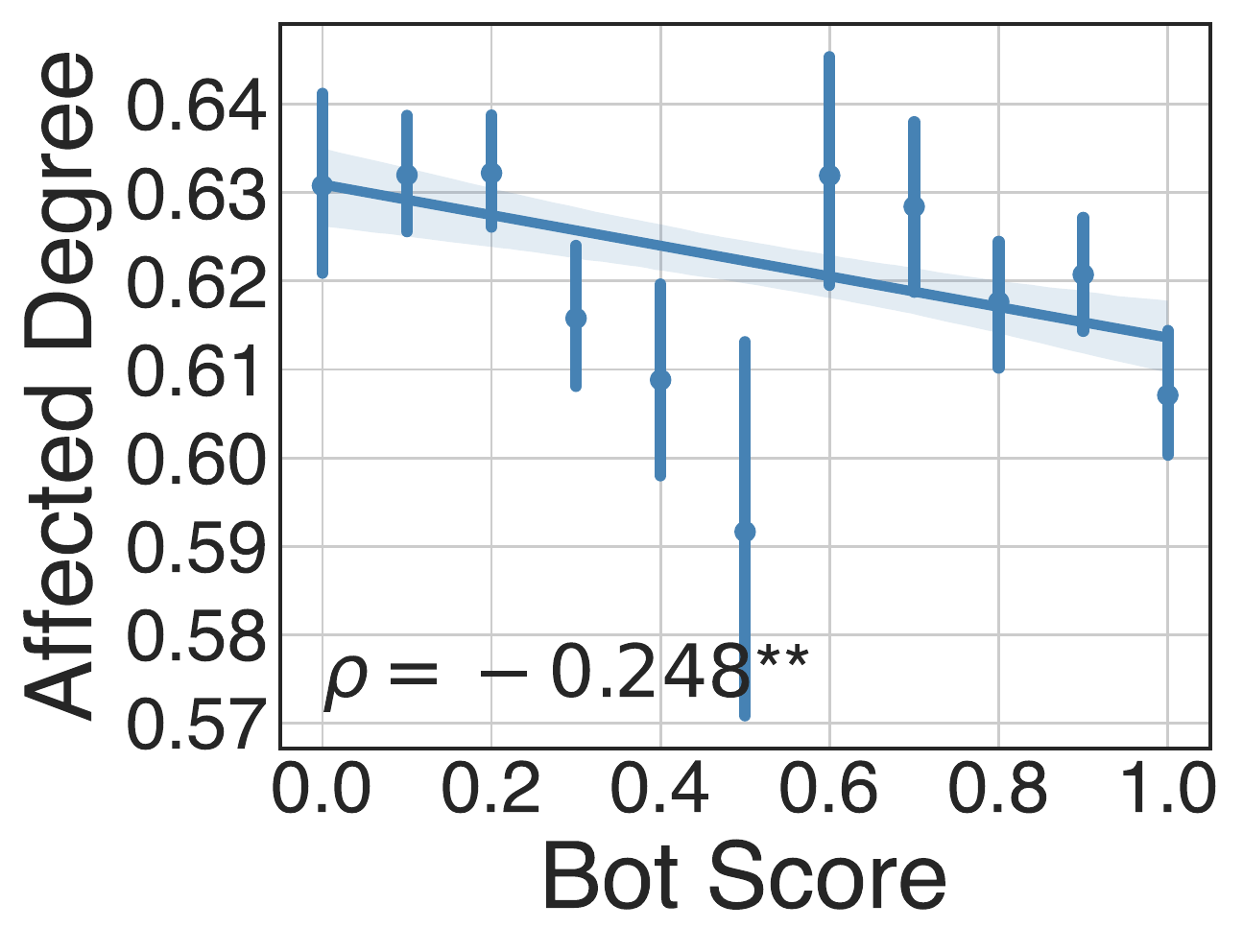}
        \includegraphics[width=0.32\columnwidth]{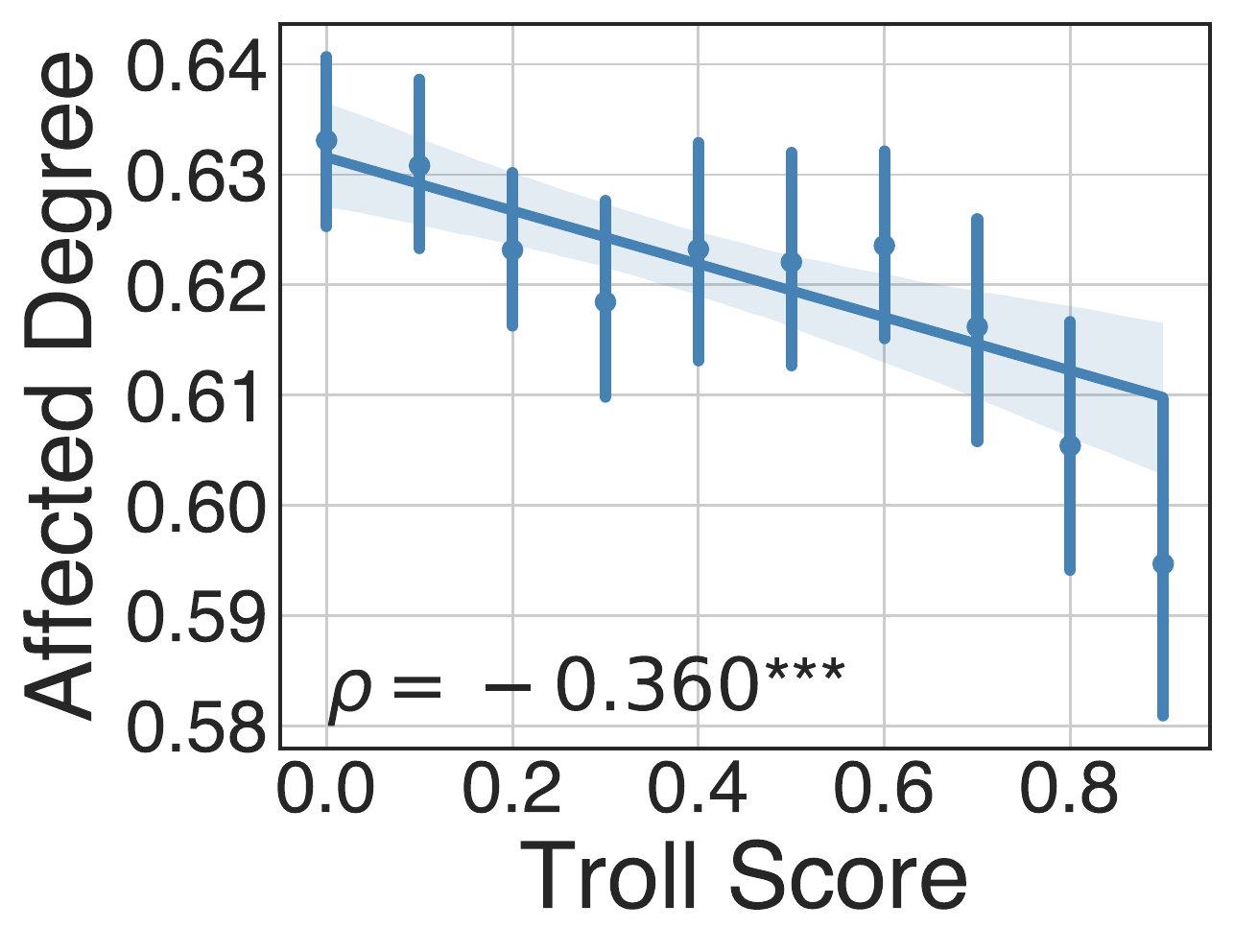}
        \includegraphics[width=0.32\columnwidth]{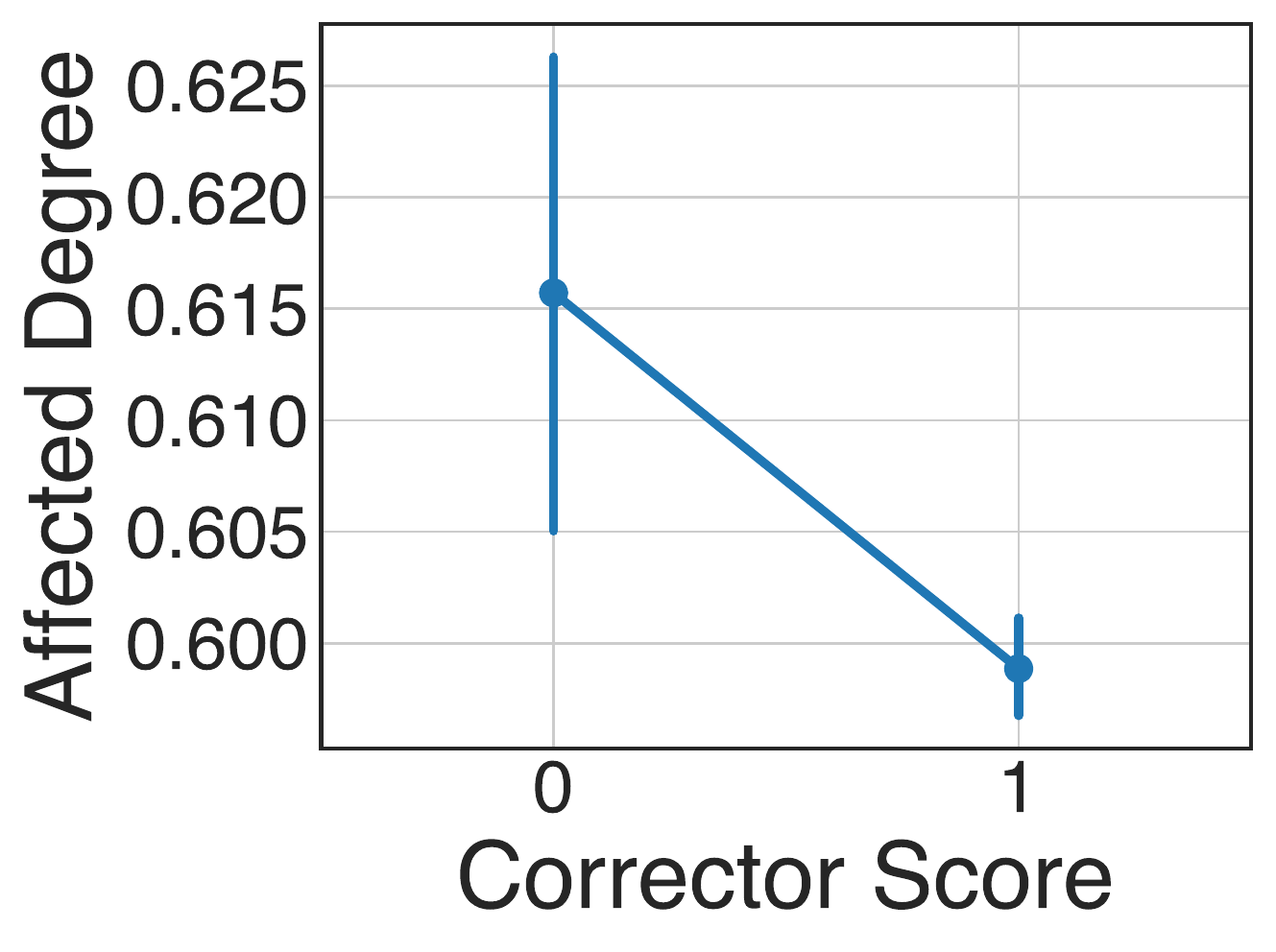}}
    \caption{Relation between Affected Degree and (L) Bot Score, (M) Troll Score, and (R) Corrector Score. $\rho$: Spearman's Correlation Coefficient. ***: $p<0.001$; **: $p<0.01$; and *: $p<0.05$.}
    \label{fig:bot_troll_corrector_vs_affected_degree}
\end{minipage}\qquad
\begin{minipage}{.48\textwidth}
    \centering
    \subfigure[MM-COVID]{
    \includegraphics[width=0.46\columnwidth]{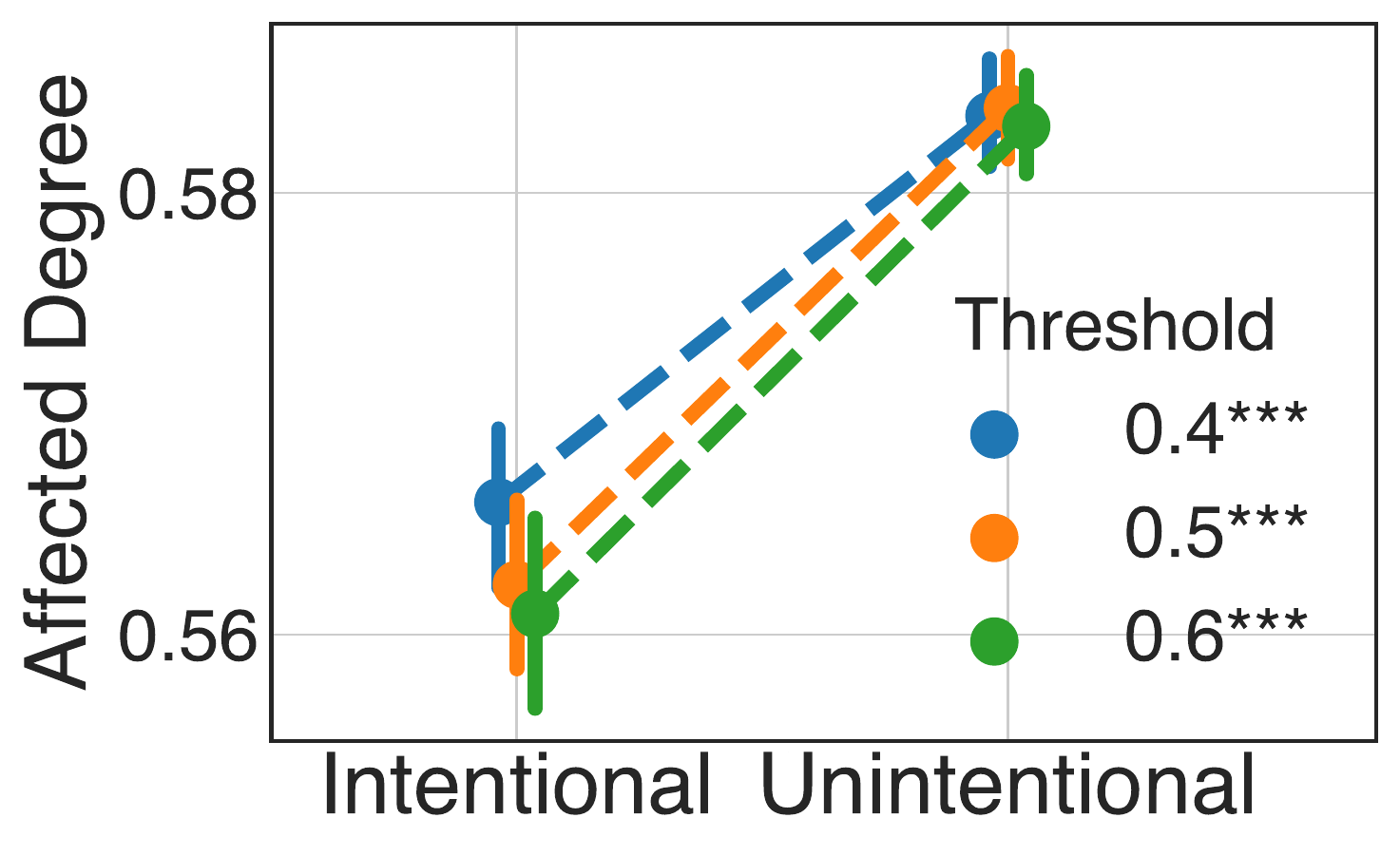}
    \includegraphics[width=0.46\columnwidth]{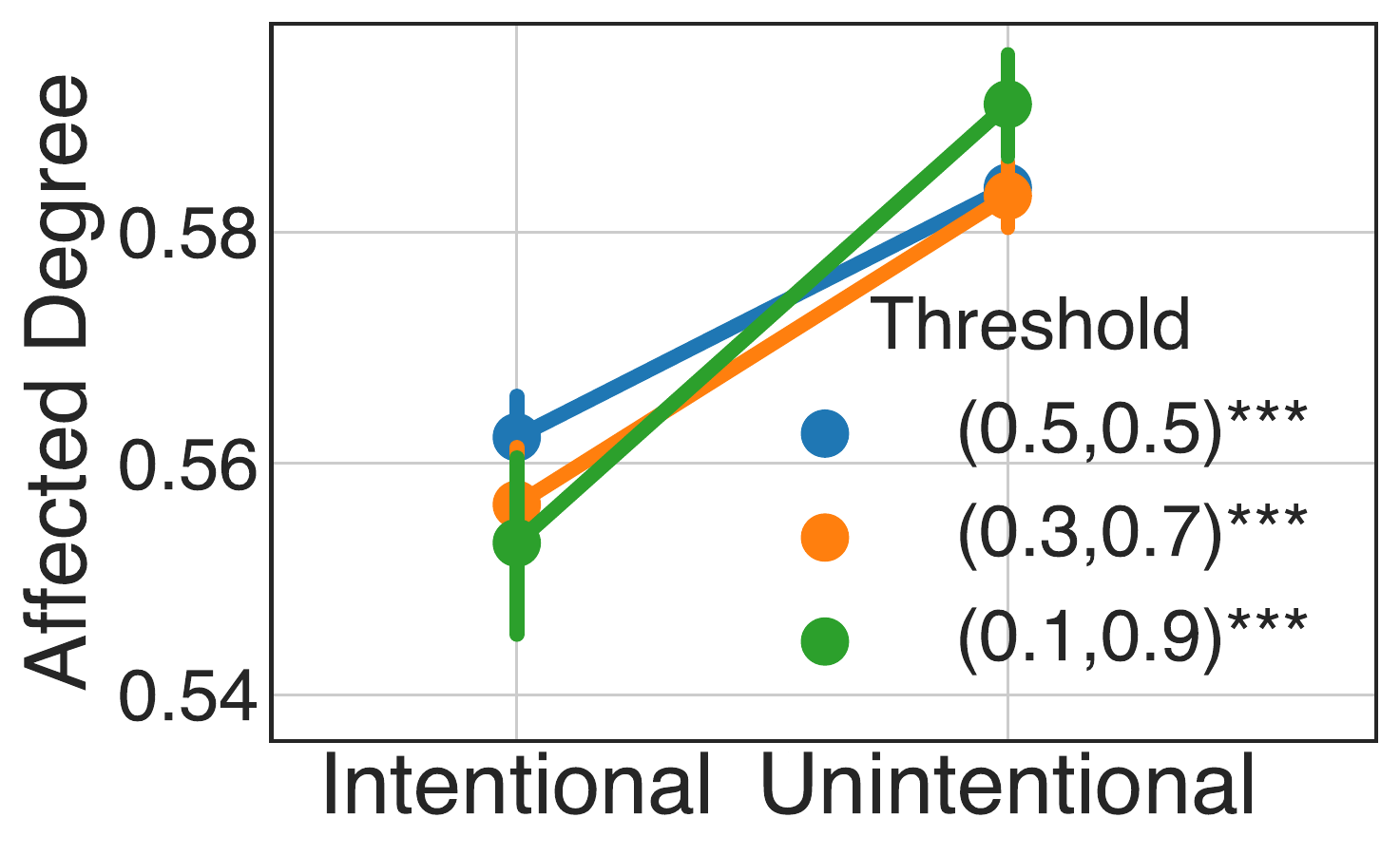}}
    \subfigure[ReCOVery]{
    \includegraphics[width=0.46\columnwidth]{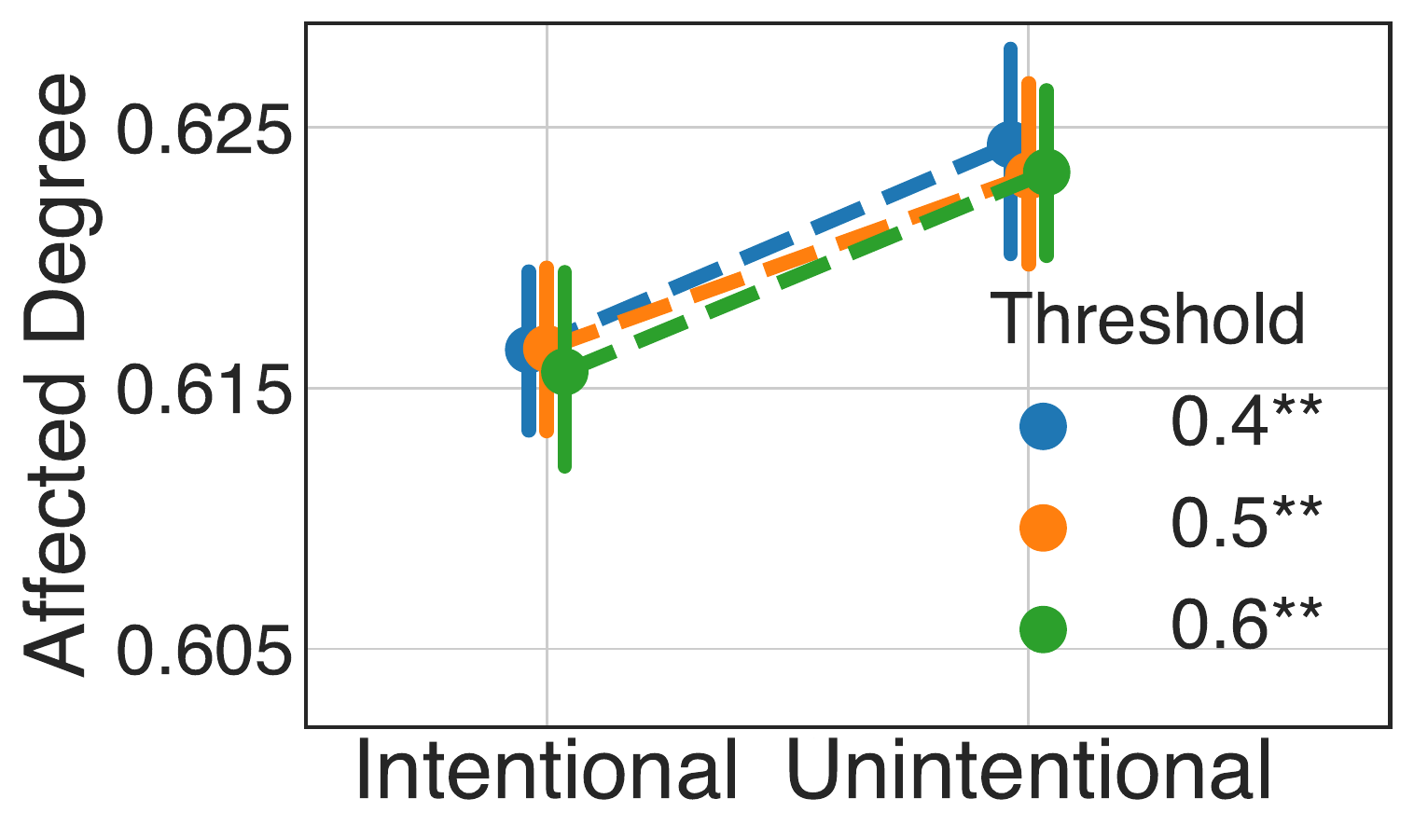}
    \includegraphics[width=0.46\columnwidth]{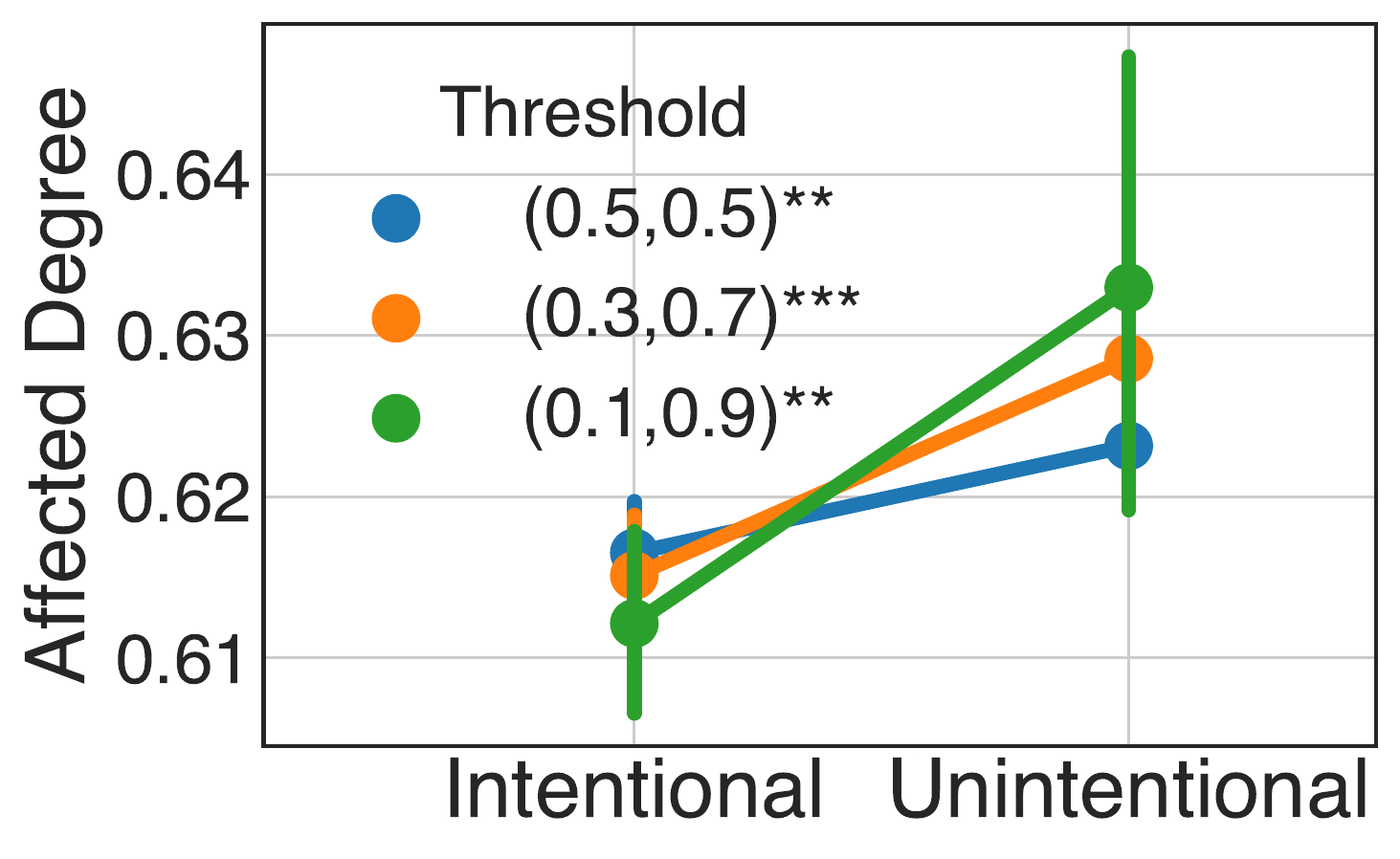}}
    \caption{Method Performance with Various Thresholds (***: $p<0.001$; **: $p<0.01$; and *: $p<0.05$)}
    \label{fig:threshold_performance}
\end{minipage}
\end{figure*}

\subsection{Experimental Results}
\label{subsec:evaluation}

With annotated intent (\textit{intentional} or \textit{unintentional}) of fake news spreaders, we verify if the assessed intent (i.e., affected degree) differs between intentional and unintentional fake news spreaders and if such difference is statistically significant. In particular, our assessed intent can be validated if affected degrees of intentional fake news spreaders are significantly less than that of unintentional fake news spreaders, i.e., if we estimate fake news spreaders who are annotated as unintentional to be more unintentional than those who are annotated as intentional.

As specified in last section, annotations are conducted at both tweet and user levels. Correspondingly, affected degrees are computed at two levels; we further obtain the user-level affected degree by averaging the affected degree of the user's posts sharing fake news. Here we present tweet-level verification results; results at the two levels reveal the same pattern, from which we can draw the same conclusions.

First, we present the distribution of affected degrees for intentional and unintentional fake news spreaders (see Fig.~\ref{fig:influence_vs_intentional_cdf}). We observe that, in general, the affected degree of intentional fake news spreaders is less than that of unintentional fake news spreaders. Specifically, the average \textit{normalized} affected degree of intentional fake news spreaders are 0.55 with MM-COVID data and 0.61 with ReCOVery data. For unintentional fake news spreaders, the value is 0.58 and 0.62, respectively. Such difference is statistically significant with a $p$-value of $\ll0.001$ on MM-COVID and $<0.01$ on ReCOVery using $t$-test. Therefore, the results validate our assessment.
We conduct the same experiment on the subset of data annotated by humans, where we can draw the same conclusion.

Second, we compare the affected degree of bots, trolls, and correctors, which all are annotated as intentional fake news spreaders, with that of others, which are annotated as unintentional fake news spreaders. The results are shown in Fig.~\ref{fig:influence_vs_intentional_box}. The results indicate that bots, trolls, and correctors all have a lower affected degree compared to unintentional fake news spreaders. The results are statistically significant with a $p$-value of $\ll0.001$ on MM-COVID and $<0.01$ on ReCOVery using ANOVA test. Meanwhile, Fig.~\ref{fig:bot_troll_corrector_vs_affected_degree} presents the relationship between affected degree and (i) bot score, (ii) troll score, and (iii) corrector score. The results reveal the same pattern: affected degree drops with an increasing bot, troll, or corrector score. In particular, both bot and troll scores are negatively correlated with affected degrees, with a Spearman's correlation coefficient $\rho\in[-0.32, -0.24]$ for bots and $\rho\in[-0.58, -0.36]$ for trolls. Results, again, validate our proposed method. Note that when investigating the relationship between affected degree and, e.g., bot score, we remove trolls and correctors to reduce noise. 

Third, we assess the result robustness. As mentioned before, a fake news spreader is labeled as an unintentional spreader with a bot (troll, or corrector) score less than a threshold value (i.e., $X\in[0,\theta), X=\{b,r,c\}$); otherwise, he or she is an intentional spreader (i.e., $X\in[\theta,1], X=\{b,r,c\}$). Varying $\theta$ among $0.4, 0.5, 0.6$, we compare again the affected degree of intentional and unintentional fake news spreaders. Results are presented in Fig.~\ref{fig:threshold_performance} (the left column). We observe that slightly adjusting the threshold value does not change our observations and conclusions made in the first experiment (i.e., the result is robust).

We lastly evaluate the proposed method as follows: we label a fake news spreader whose $X\in[0,\theta)$ as an unintentional spreader, and whose $X\in[1-\theta,1]$ as an intentional spreader. By decreasing $\theta_X$, a fake news spreader is required to have a lower bot (troll, or corrector) score to be unintentional and a higher bot (troll, or corrector) score to be intentional. In other words, a smaller $\theta$ corresponds to a more strict annotation (\textit{intentional} or \textit{unintentional}) of fake news spreaders. We vary $\theta$ among $0.5, 0.3, 0.1$ -- correspondingly, $1-\theta$ varies among $0.5, 0.7, 0.9$ -- and compare the affected degree of intentional and unintentional fake news spreaders. Results are presented in Fig.~\ref{fig:threshold_performance} (the right column). We observe that the affected degree of intentional fake news spreaders is always less than that of unintentional fake news spreaders with various thresholds. More importantly, such pattern becomes more significant with a smaller $\theta$ (i.e., a more strict annotation), which validates the effectiveness of our assessment.

Finally, we point out that we experiment with (i) external affected degree, (ii) internal affected degree, (iii) combined (external+internal) affected degree, and (iv) combined affected degree where the external one merely exists between post pairs sharing the same news. The combined one (i.e., iii) is the one where significant and consistent patterns are discovered on both datasets.

\begin{table*}[t]
\begin{center}
\begin{minipage}{.27\textwidth}
\centering
\includegraphics[width=\columnwidth]{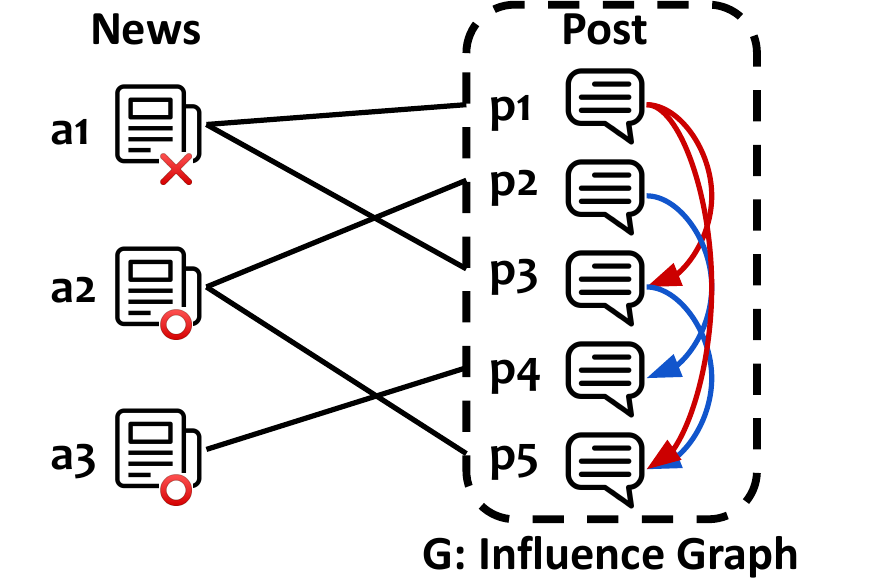}
\captionof{figure}{News-post Graph}
\label{fig:het_graph}
\end{minipage}\qquad
\begin{minipage}{.63\textwidth}
\centering
\captionof{table}{Method Performance (Using AUC Scores) with Heterogeneous Graph Neural Networks (HetGNN) in Fake News Detection}
\label{tab:deep_learning}
\begin{tabular}{lrrrrrrrr} \toprule[1pt]
\multicolumn{1}{l}{} & \multicolumn{4}{c}{\textbf{MM-COVID}} & \multicolumn{4}{c}{\textbf{ReCOVery}} \\ \midrule[0.5pt]
{\begin{tabular}[c]{@{}l@{}}\textbf{\% Labeled News}\end{tabular}} &  20\% & 40\% & 60\% & 80\% & 20\% & 40\% & 60\% & 80\% \\ 
 $G_{\textsc{Random}}$
                                    & 0.829          & 0.856          & 0.876          & 0.902 
                                    & 0.647          & 0.654          & 0.660          & 0.674  \\ 
$G_{\textsc{Subgraph}}$             & 0.817          & 0.861          & 0.890          & \textbf{0.915}
                                    & 0.820          & 0.845          & 0.869          & \textbf{0.908} \\  
$G$                                 & \textbf{0.869} & \textbf{0.864} & \textbf{0.902} & 0.905 
                                    & \textbf{0.825} & \textbf{0.863} & \textbf{0.883} & 0.881 \\ \bottomrule[1pt]
\end{tabular}
\end{minipage}
\end{center}
\end{table*}

\section{Utilizing Intent of News Spreaders to Combat Fake News}
\label{sec:method_application}

Using MM-COVID and ReCOVery data, we evaluate the effectiveness of user intent in news propagation to detect fake news. We first employ the assessed affected degree of posts in news propagation within a traditional machine learning framework. Then, we utilize the proposed influence graph within a deep learning framework. 

\begin{table}[t]
\centering
\caption{Method Performance with Hand-crafted Features in Fake News Detection.  Here, $K$: the first (earliest) $K$ posts spreading the news available for news representation; Ranking: feature importance ranking of affected degree of posts in the prediction model.}
\label{tab:statistical_learning}
\begin{tabular}{cccc} \toprule[1pt]
 & \textbf{K} & \textbf{AUC Score} & \textbf{\begin{tabular}[c]{@{}r@{}} Ranking\end{tabular}} \\ \midrule[0.5pt]
 & 10 & 0.918 ($\pm$0.009) & 2 \\
\textbf{} & 20 & 0.912 ($\pm$0.015) & 2 \\
\textbf{MM-COVID} & 30 & 0.927 ($\pm$0.021) & 2 \\
\textbf{} & 40 & 0.923 ($\pm$0.012) & 2 \\
\textbf{} & All & 0.935 ($\pm$0.005) & 3 \\ \midrule[0.5pt]
 & 10 & 0.891 ($\pm$0.007) & 5 \\
\textbf{} & 20 & 0.898 ($\pm$0.007) & 3 \\
\textbf{ReCOVery} & 30 & 0.903 ($\pm$0.004) & 3 \\
\textbf{} & 40 & 0.909 ($\pm$0.014) & 4 \\
\textbf{} & All & 0.925 ($\pm$0.009) & 5 \\ \bottomrule[1pt]
\end{tabular}
\end{table}

\vspace{1mm}
\noindent\textbf{I. Combating Fake News by Affected Degree.}
For each news article, we manually extract over 100 (propagation and content) features as its representation. Propagation features include the average (internal, external, and combined) affected degree of posts spreading the news and a set of widely-accepted propagation features. Content features are extracted using LIWC~\cite{pennebaker2015development}. See Appendix \ref{apx:detection_implementation_details} for feature details. Five-fold cross-validation and XGBoost~\cite{chen2016xgboost} are then used with these features for training and classifying news articles. Results indicate that this method correctly identifies fake news with an AUC score of around 0.93. As a comparison, dEFEND~\cite{shu2019defend}, a state-of-the-art method that detects fake news by news content and propagation information, performs around 0.90. Furthermore, we observe that, as presented in Tab.~\ref{tab:statistical_learning}, the proposed method performs above 0.89 with limited propagation information of news articles, i.e., at an early stage of news dissemination on social media.
Notably, internal affected degree of posts greatly contributes to detecting fake news, whose feature importance assessed by XGBoost ranks top five all along.

\vspace{1mm}
\noindent\textbf{II. Combating Fake News by Influence Graph.}
We construct the news-post heterogeneous graph (shown in Fig.~\ref{fig:het_graph}); a post is connected with a news article if the post shares the news, and the relation among posts is modeled by the proposed influence graph $G$. Then, we train the HetGNN (Heterogeneous Graph Neural Network) model~\cite{zhang2019heterogeneous} with this news-post graph to learn news representation, with which XBGoost~\cite{chen2016xgboost} is further utilized to predict fake news. Varying the percentage of labeled news from 20\% to 80\%, this method performs with an AUC score ranging from 0.83 (with small-scale training data) to 0.91 (with relatively large-scale training data) on two datasets. To further evaluate the proposed influence graph $G$, we consider two variant groups of the constructed heterogeneous graph as baselines. One replaces $G$ by a random version ($G_{\textsc{Random}}$): Based on our graph sparsification strategy (see Appendix \ref{apx:graph_sparsification}),  we construct the random graph by randomly selecting a hundred posts for each post ensuring that no self-loops are formed in this graph. The other replaces $G$ by its subgraph (i) with internal influence only ($G_{\textsc{Internal}}$); (ii) with external influence only ($G_{\textsc{External}}$); or (iii) with internal and external influence but the latter only exists between two posts sharing the same news ($G_{\textsc{Same News}}$). Tab.~\ref{tab:deep_learning} presents the full result; $G_{\textsc{Subgraph}}$ in the table refers to $G_{\textsc{Same News}}$, which performs best among all subgraphs. We observe that in general, the proposed influence graph outperforms its variants in detecting fake news, especially with limited training data. See Appendix \ref{apx:detection_implementation_details} for other implementation details.

\section{Conclusion and Future Work}
\label{sec:conclusion}

We look into the phenomenon that social media users can spread fake news unintentionally. With social science foundations, we propose influence graph, with which we assess the degree to which fake news spreaders are unintentional (denoted as \textit{affected degree}). Strategies to sparse the influence graph and normalize the affected degree by determining its upper bound are presented as well. We develop manual and automatic annotation mechanisms to obtain the ground-truth intent (\textit{intentional} or \textit{unintentional}) of fake news spreaders for MM-COVID and ReCOVery data. We observe that the affected degree of intentional fake news spreaders are significantly less than that of unintentional ones, which validates our assessments. 
This work helps combat fake news from two perspectives. First, our assessed intent helps determine the necessity of a fake news spreader being nudged or recommended with (users active in sharing) facts. Second, we present that the assessed spreader intent and proposed influence graph effectively help detect fake news with an AUC score of around 0.9. 

\vspace{1mm}
\noindent \textbf{Limitations and Future Work}: We effectively assess the degree to which fake news spreaders are unintentional, but remain the task to \textit{classify} a fake news spreader as an intentional or unintentional spreader. We point out that merely relying on determining a threshold for affected degree is barely enough. To address this problem, we aim to propose a more complicated classification model in the near future, which involves non-posting behavior (e.g., commenting, liking, and following) of news spreaders.

\section*{Acknowledgments}
This research was supported in part by the National Science Foundation under award CAREER IIS-1942929. We sincerely appreciate the positive and constructive comments of the reviewers. We also thank Chang Liu, Shengmin Jin, and Hao Tian for their useful suggestions in data annotation.

\balance 

\bibliographystyle{ACM-Reference-Format}
\bibliography{references}  

\newpage
\begin{appendix}

\section{Sparsification of Influence Graph}
\label{apx:graph_sparsification}

Influence graph can be a \textit{tournament} in the worst case, taking much space. To sparsify the graph, we add one more constraint in the graph construction:  $(p_i, p_j)\in E$ if $\Delta t_{ij} \leq \theta_t$. Thus, we assume that each node (post) can be connected with (affected by) at most $\theta_t$ previous nodes (posts), which can be viewed as an extension of the Markov property. We vary $\theta_t$ in \{1, 10, 100, 1000\} and ultimately set $\theta_t=100$ as all experimental results converge at this point.

\section{Data Statistics}
\label{apx:data_statistics}

Tab.~\ref{tab:data_statistics} shows the statistics of MM-COVID and ReCOVery datasets.

\begin{table}[t]
    \centering
    \caption{Data Statistics}
    \label{tab:data_statistics}
    \subtable[\small{on News Credibility}]{\label{subtab:news_credibility}
    \begin{tabular}{llrr} 
    &  & \textbf{\begin{tabular}[c]{@{}r@{}}MM-\\COVID\end{tabular}} & \textbf{\begin{tabular}[c]{@{}r@{}}Re-\\COVery\end{tabular}} \\ \midrule[1pt]
    \textbf{\# News} & \textbf{Fake} & 355 & 535 \\ 
    \multicolumn{1}{l}{} & \textbf{True} & 448 & 1,231 \\ 
    \textbf{\# Tweets} & \textbf{Sharing Fake News} & 16,500 & 26,657 \\ 
    \multicolumn{1}{l}{} & \textbf{Sharing True News} & 20,905 & 117,087 \\\bottomrule[1pt]
    \end{tabular}}\qquad
    \subtable[\small{on Intent of Fake News Spreaders}]{\label{subtab:user_intention}
    \begin{tabular}{clrr} 
    \multicolumn{1}{l}{} & \textbf{} & \textbf{\begin{tabular}[c]{@{}r@{}}MM-\\COVID\end{tabular}} & \textbf{\begin{tabular}[c]{@{}r@{}}Re-\\COVery\end{tabular}} \\ \midrule[1pt]

    \multirow{5}{*}{\rotatebox{90}{\begin{tabular}[c]{@{}c@{}}\textbf{\# Fake} \\ \textbf{News} \\ \textbf{Spreaders}\end{tabular}}}
     & \textbf{Unintentional} & 9,237 & 7,911 \\ 
     & \textbf{Intentional} & 4,285 & 7,327 \\ 
     & \textbf{Bots} & 3,195 & 6,266 \\ 
     & \textbf{Trolls} & 1,024 & 2,687 \\ 
     & \textbf{Correctors} & 463 & 6 \\ \midrule[0.5pt]
    
     \multirow{5}{*}{\rotatebox{90}{\begin{tabular}[c]{@{}c@{}}\textbf{\# Tweets} \\ \textbf{Sharing} \\ \textbf{Fake News} \end{tabular}}} & \textbf{\begin{tabular}[c]{@{}l@{}}by Unintentional\end{tabular}} & 10,519 & 10,733 \\ 
     & \textbf{\begin{tabular}[c]{@{}l@{}}by Intentional\end{tabular}} & 5,953 & 12,502 \\ 
     & \textbf{by Bots} & 4,530 & 11,035 \\ 
     & \textbf{by Trolls} & 1,360 & 4,240 \\ 
     & \textbf{by Correctors} & 789 & 8 \\ \bottomrule[1pt]
    \end{tabular}}
\end{table}

\section{Reproducibility Details in Fake News Detection}
\label{apx:detection_implementation_details}

We have 109 hand-crafted (linguistic and propagation) features. Propagation features include the average external, internal, and combined affected degree of posts sharing the news; the average sentiment score (assessed by $\mathsf{flair}$~\cite{akbik2019flair})\footnotemark and the average number of reposts, favorites, hashtags, mentions, symbols, quotes, and replies of posts sharing the news; and the average number of followers, friends, favorites, list memberships, and status updates of users spreading the news. Content features include all that can be extracted by LIWC~\cite{pennebaker2015development}, each of which falls into one of the categories including word count, summary language variables, linguistic dimensions, other grammars, and psychological processes.

With HetGNN, we use pre-trained transformers to extract content features of nodes (Longformer~\cite{beltagy2020longformer} for news stories and Sentence-BERT~\cite{reimers2019sentence} for tweets). The news node is associated with the news embedding and the average embedding of its connected posts. The post node is associated with the post embedding, the average embedding of its connected news, and the average embedding of its connected posts. Hence, the Bi-LSTM length of news content encoder is two, and that of post content encoder is three. For both datasets, the embedding dimension of HetGNN is 1024, the size of sampled neighbors set for each node is 23 (3 news nodes plus 20 post nodes), the learning rate is 0.0001, and the maximum number of training iterations is 50. The other hyperparameters are set the same as mentioned in \cite{zhang2019heterogeneous}.

\footnotetext{\url{https://github.com/flairNLP/flair}}

\end{appendix}

\end{document}